\documentclass[12pt]{book}
\usepackage{amsmath}
\usepackage{amssymb}
\usepackage{graphicx}
\usepackage{subfigure}
\usepackage{epstopdf}
\usepackage{achemso}
\usepackage{color}
\usepackage[ansinew]{inputenc}
\usepackage{setspace}

\def\la{\langle}
\def\ra{\rangle}

\def\rv{{\mathbf{r}}}
\def\tv{{\mathbf{t}}}

\def\fh{\mathcal{F}_h}
\def\fl{\mathcal{F}_l}

\def\ch{\mathaccent 94}
\def\til{\mathaccent "7E }

\def \r0{\ch{\rho}^{(0)}}

\def\q0{q_{0}}

\def\lb{l_{B}}
\def\lB{l_{B}}
\def\l0{l_{0}}
\def\lo{l_{{OSF}}}

\def\lD{\lambda_D}

\renewcommand{\chaptermark}[1]{\markboth{\MakeUppercase{\chaptername\ \thechapter.\ \ \ PE and PA effects...}}{}}
\doublespace

\begin{document}
\setcounter{chapter}{3}
\chapter{Polyelectrolyte and polyampholyte effects in synthetic and biological macromolecules}
{\Large Ngo Minh Toan, Bae-Yeun Ha and D. Thirumalai}

\newpage

\noindent{\Large\bf Chapter 4: Polyelectrolyte and polyampholyte effects in synthetic and biological macromolecules}\\
\ \\
{\large Ngo Minh Toan$^1$, Bae-Yeun Ha$^2$, and D. Thirumalai$^{1,3}$}\\
\ \\
$^1$  Biophysics Program, Institute for Physical Science and Technology, University of Maryland at College Park, College Park, Maryland, USA, 20742. Tel.: +1 301-405-7568. Email: tmngo@umd.edu\\
$^2$  Department of Physics and Astronomy, University of Waterloo, Waterloo, Ontario, Canada N2L 3G1. Tel. +1 (519) 888-4567 ext. 37004. Email: byha@uwaterloo.ca\\
$^{1,3}$  Department of Chemistry and Biochemistry, University of Maryland at College Park, College Park, Maryland, USA, 20742. Tel.: +1 301-405-4803. Fax: +1 301-314-9404. Email: thirum@umd.edu

\newpage
\section*{Abstract}
The nature of electrostatic interactions involving polyanions modulate the properties of both synthetic and biological macromolecules. Although intensely studied for decades the interplay of many length scales has prevented a complete resolution of some of the basic questions such as how the electrostatic persistence length ($l_e$) varies with ionic strength ($I$). In this review we describe certain characteristics of  polyelectrolytes (PAs) and polyampholytes (PAs), which are polymers whose monomers have a random distribution of opposite charges. After reviewing the current theoretical understanding of the dependence of $l_e$ on $I$ we present experimental data that conform to two distinct behavior. For RNA and DNA it is found that that $l_e \sim I^{-1}$ whereas for some proteins and other polyelectrolytes $l_e \sim i^{-\frac{1}{2}}$. A scaling type theory, which delineates charge correlation and pure polyelectrolyte effects for the shape of PAs that is valid over a wide range of salt concentration is described. We also use theory and simulations to argue that the distinct stages in the kinetics of collapse of PAs (with a net charge that is small enough to induce globule formation) and PEs (relevant for RNA folding) are similar. In both cases the major initial conformation change involves formation of metastable pearl-necklace structures. In the coarsening process large clusters (pearls) grow at the expense of smaller ones by a process that is reminiscent of Lifshitz-Slyozov mechanism.  Finally, recent theories and single molecule experiments on stretching of single stranded DNA and PEs further sheds insights into the complex behavior of charged macromolecules.  The survey, which is limited to very few topics, shows the importance of polyelectrolyte effects in a wide range of disciplines.

\newpage

\section{Introduction}
The charged nature of biological macromolecules (not limited to DNA, RNA, and proteins) and synthetic polymers makes the study of polyelectrolyte (PE) properties important. Shape fluctuations in charged biomolecules control a number of functions such as transcription, packaging of DNA and RNA into phage heads as well as interactions involving nucleic acids with other biological macromolecules (higher order structural organization of DNA upon interaction with histones for example). An example of functional significance is the spectacular phenomenon of reversible condensation of DNA into toroidal structure in the presence of multivalent cations. Upon condensation the volume occupied by DNA decreases by nearly four or more orders of magnitude. Description of counterion-induced condensation of DNA and synthetic PEs requires accounting for density fluctuations associated with counterions and their coupling to conformations of the macromolecule. Similarly, the phase diagram of charged synthetic polymers, especially in aqueous solution, is complicated and is critically dependent on ion concentration as well their size and shapes. The behavior of PEs is governed by a number of factors such as the intrinsic properties of the backbone (extent of flexibility), length scale associated with ion fluctuations, and even the chemical nature of the counterions. As a result it has been difficult to develop a coherent theory that can capture all the properties of polyelectrolytes.

In this Chapter we confine ourselves predominantly to the study  of properties of an isolated PE and not so unrelated problem of polyampholytes (PAs) and develop concepts that have been particularly useful in understanding a number of aspects of RNA folding and a few problems associated with DNA.  One focus, described in Section 2,  is to elucidate the current understanding of how the persistence of charged chains, a basic property of polymers, depends on salt concentration. Although considerable work has been done in this area (see an excellent review~\cite{BJ_Rev} by Barrat and Joanny that still remains relevant) a clean solution to this problem has not been obtained. This is probably because there are multiple scenarios that depend on the molecular weight of the polymer and the concentration of chemical characteristics of the counterions.  In Sections 3 and 4 we provide a simple picture of shapes of polyampholytes (PAs), which are polymers with random distribution of positively and negatively charged monomers. In Section 5 we describe the response of PEs to mechanical force, which have provided insights not only into the problem of electrostatic contribution to persistence length of charged polymers but also have been instrumental in describing length scale-dependent elasticity of single and double stranded DNA.  We describe  simulations (Section 6) and theory (Section 7) of collapse of PAs and PEs, which prove to be important in understanding folding of RNA. In Section 5 we describe the response of PEs to mechanical force, which have provided insights not only into the problem of electrostatic contribution to persistence length of charged polymers but also have been instrumental in describing length scale-dependent elasticity of single and double stranded DNA. The Chapter ends with concluding remarks in Section 8.

\section{Persistence length of polyelectrolytes}
The problem of electrostatic persistence length of polyelectrolytes has been studied for more than three decades using a variety of methods, including analytical treatment, experiments and simulations. Due to an interplay of a  number of length scales, compared to the case of neutral polymers, a complete theoretical understanding of persistence length changes in PE as a function of salt-concentration is difficult. The electrostatic persistence length is defined as the contribution to the chain total persistence length in addition  to the ``bare'' (intrinsic) value by electrostatic interactions among the charged monomers. Theoretical description of the characteristics of an isolated PE chain began with two pioneering and independent  studies by Odijk and by Skolnick and Fixman (hereafter referred to as OSF theory), which addressed the problem of electrostatic contribution to persistence length of an intrinsically stiff chain. Besides the usual chain intrinsic stiffness of a  neutral chain, which is characterized by bending rigidity (persistence length, Kuhn length, monomer length, interaction strength), electrostatic interactions among monomers can further stiffen the chain. The interplay of the chain intrinsic stiffness and the electrostatic interactions, which are usually modeled by screened Coulomb potential using Debye-H\"uckel theory, potentially gives rise  to a number of scenarios, which require a number of approaches.

\def\half{{1\over 2}}

\subsection{Intrinsically stiff chains: The Odijk-Skolnick-Fixman theory}
Odijk~\cite{Odijk1977} and, independently, Skolnick and Fixman~\cite{SF1977} were the first to develop a theory for persistence length of polyelectrolyte. Following  Barrat and Joanny~\cite{BJ_Rev}  we model a polyelectrolyte as a worm-like chain with a contour length $L$ and ``bare'' persistence length $l_0$, and that carries charges separated by a distance $A$ along its contour. The electrostatic interaction between two charges separated at a spatial distance $r$ is modeled by the Debye-H\"uckel potential:
\begin{eqnarray}
v_{DH}(r) = k_BT\lb {\exp\left(-{r\over \lD}\right)\over r}
\end{eqnarray}
where $\lb = e^2/(\varepsilon k_BT)$ (in Gaussian units as in Refs.~\cite{Joanny91,Rubinstein04}) is the Bjerrum length and is about 0.7nm at room temperature, and $\lD$ (or, interchangeably, $\kappa^{-1}$) is the Debye screening length, which is given by
\begin{eqnarray}
\kappa^{-1} = \lD = {1\over \sqrt{4\pi \lb I}}
\end{eqnarray}
with $I$ being the ionic strength of the solution.

For simplicity we confine the chain in two dimensions so it is a planar curve. The results in three dimensions are similar. The chain configuration is characterized by the angle function (of arc-length $s$) $\theta(s)={\tv(s)\dot\tv(0)}$ where $-L/2 \le s \le L/2$ and $\tv(s)$ is the tangent vector at $s$. We assume that the chain is almost linear (strong charge limit) or $\theta(s)$ is small, so that the electrostatic interaction between two charges at $s1$ and $s2$ can be expanded around the rod-like conformation, $v_{DH}(|s_2-s_1|)$. With these conditions the chain Hamiltonian $H(\theta)$ can be written for a given configuration $\{\theta(s)\}$ as:
\begin{eqnarray}\label{eq:PE_Hamiltonian}
H[\theta] = H_0 + {1 \over 2} k_B T \int_{-L/2}^{L/2}ds  \int_{-L/2}^{L/2}ds'
{d\theta(s) \over ds} \left[ \l0 \delta(s-s') + K(s,s')\right]
{d\theta(s') \over ds}
\end{eqnarray}
Here, $H_0$ is the electrostatic energy of a rod, the term proportional to $l_0$ is the usual bending energy of the worm-like chain that is proportional to the square of radius of curvature, and  the electrostatic interactions described using the kernel $K(s,s')$ which, for $s>s'$, is
\begin{eqnarray}\label{eq:K_ss}
K(s,s')= {1 \over A^2} \int_{-L/2}^{s'}ds_1\int_{s}^{L/2} ds_2
 v_{DH}'(s_2-s_1) {(s_2-s) (s'-s_1)\over (s_2-s_1)}
\end{eqnarray}
with $v_{DH}'(s) = {dv_{DH}\over ds}$. The result has been obtained with the only assumption that, within $\lD$ the chain remains in an almost rod-like configuration. For very long chains, $L/\lD \gg 1$, we can extend the limits in the integrals of \eqref{eq:K_ss} to infinity and $K(s,s')$ becomes a functions of $s-s'$,
\begin{eqnarray}\label{eq:K_s}
  K(s) =  {1\over 6 A^2} \int_0^\infty dx\ {x^3 \over x+s}
v_{DH}'(x+s)
\end{eqnarray}
which has the Fourier transform
\begin{eqnarray}\label{Kq}
\til{K}(q) & = &\int_0^{\infty} ds \exp(i q s) K(s) \cr
     & = & \lo {2 \kappa^2  \over  q^2} \left(  {\kappa^2+q^2 \over q^2}
 \ln\left({\kappa^2+q^2 \over \kappa^2}\right)
-1  \right)
\end{eqnarray}
where
\begin{eqnarray}\label{lpe}
\lo =\lb / (4 A^2 \kappa^2)
\end{eqnarray}
is called the celebrated Odijk-Skolnick-Fixman (OSF) length. The statistical properties of the chain are obtained by integrating the Boltzmann factor $\exp\left(H([\theta]/k_BT\right)$ over all possible configurations, i.e. over all functions $\theta(s)$. Since the energy is a quadratic function of $\theta$, the integration can be carried out analytically. For instance, we can calculate the mean squared angle $\la \theta(s)^2 \ra$ between the chain directions at the origin and at arc-length $s$. This quantity characterizes the local flexibility of the chain. With the approximate Hamiltonian~\eqref{eq:PE_Hamiltonian}, we have
\begin{eqnarray}\label{th2}
\la\theta(s)^2 \ra = {4\over \pi}
 \int_0^\infty
 dq\  { {\rm sin}^2(qs/2)  \over q^2} \ {1 \over
\l0 + \til{K}(q) }
\end{eqnarray}
For a neutral chain, $\til{K}(q) = 0$ in Eq.~\eqref{th2} and we have
\begin{eqnarray}\label{eq:theta2_neutral}
\la\theta(s)^2\ra = {s \over \l0}
\end{eqnarray}
which varies linearly with $s$, and the proportional constant is the inverse of the persistence length.

For a charged chain, $\la\theta(s)^2\ra$ cannot be analytically calculated. Thus, to describe the expected scaling behavior, we work in the limits of small and large $s$. At large $s$
\begin{eqnarray}\label{odijk}
\la\theta(s)^2\ra = {s \over \l0+\lo}
\end{eqnarray}
which expresses the fact that at long length scales, the chain conformation can be described by an effective persistence length $\l0+\lo$, which is the sum of a ``bare'' and of an electrostatic contribution. It indicates that the influence of the screened electrostatic interactions can extend far beyond their range $\lD$, since in weakly screened solutions, $\lo$ is much larger than $\lD$. The electrostatic persistence length also decreases with salt concentration as $\lo\simeq I^{-1}$, whereas the Debye screening length has a slower decay $\lD\simeq I^{-1/2}$

On the other hand, for small $s$, \eqref{th2} reduces exactly to that of a neutral chain in Eq.\eqref{eq:theta2_neutral}. This means that the chain statistics at short scales are not modified by electrostatic interactions. The crossover between the ``intrinsic'' regime described by \eqref{eq:theta2_neutral} and the ``electrostatic'' regime described \eqref{odijk} takes place when the electrostatic interactions become strong enough to perturb the statistics of the neutral flexible chain. The crossover length $s_c$ can be obtained qualitatively from the following argument. If a small chain section, of length $s < \kappa^{-1}$, is bent to form an angle $\theta$, the cost in ``bare'' curvature energy is $k_B T \l0 \theta^2 /s$, while the electrostatic energy is $ k_B T \lb  (s/A)^2 (\theta^2/s) $. The two energies are comparable for $s\simeq s_c$, which gives $s_c \sim A (\l0/\lb)^{1/2}$.

The picture that emerges from this calculation is that the chain flexibility depends on the length scale. At short scales, $s<s_c$, the chain structure is determined by its  bare rigidity $\l0$, while the electrostatic rigidity \eqref{lpe} dominates at large scales. Large length scale properties, such as the radius of gyration can be determined by applying the standard formula for semi-flexible chains of persistence length $\lo$.

The only approximation required to obtain the Odijk-Skolnick-Fixman length is the expansion that yields equation \eqref{eq:PE_Hamiltonian}. The calculation is therefore consistent if the angle $\la\theta (\kappa^{-1})^2\ra$ is small compared to unity, which holds good if the bare persistence length is much larger than $l_{OSF}$. In other words, the requirement $\la\theta (\kappa^{-1})^2\ra \ll 1$ is then equivalent to $s_c \lo / (\l0 (\l0+\lo)) \simeq s_c/\l0 \ll 1$. i.e. the angular deflection of the chain must be small when the crossover region is reached. In other words, the angular fluctuations that take place before the electrostatic interactions can come into play and rigidify the chain should not be too strong.  If these fluctuations are too large, i.e. if the chain is too flexible, the perturbation expansion that underlies the OSF calculation can break down. The criterion for the validity of the calculation can be simply written, in the limit of weak screening, as $\l0 > A^2/\lb$. This implies that the OSF calculation should be directly applicable to stiff chains such as DNA ($\l0 \sim 50{\rm nm}$), but has to be reconsidered for flexible chains such as polystyrene sulfonate ($\l0 \sim 1nm$).

Multiple lines of inquiry have questioned the validity of the OSF theory (see below). All of them hinge on the assumption that when chain fluctuations are taken into account the persistence length is renormalized and the dependence on $\lD$ deviates from the predictions of OSF theory. Starting with the work of Barrat and
Joanny~\cite{Barra93}, several variational calculations have shown that,
when the chains are intrinsically flexible, the electrostatic
persistence length $l_e \sim \lD^1$. This dependence  is valid when the controlling parameter
$\frac{l_o l_B}{A^2} \sim O(1)$, i.e., when the chain is in the so-called non-asymptotic
regime. Li and Witten~\cite{Li95Macro} argued that, unlike the
variational theories, approximate inclusion of fluctuations
still leads to the result predicted by OSF even if the
backbone is flexible. In other words, the OSF result is always valid at least for long chains regardless of the backbone stiffness. In the above referenced variational theories it is generally believed that the electrostatic interactions generally stiffen the chain significantly so that the trial
Hamiltonian should consist of terms that account for the electrostatic-mediated interaction rigidity.

The controversies associated with the validity of the OSF theory were further addressed using two related theoretical studies that used different variational methods \cite{Thirum99,Netz99EurPhysJ}. Using the most general variational approach these studies showed that the OSF result is recovered whenever $\frac{l_o l_B}{A^2} \ll 1$. Thus, in both the stiff and intrinsically flexible limit the classical result is obtained provided the chain is long. Indeed, these theories supported a scaling-type argument due to Khokhlov and Khachaturian~\cite{Khokhlov82Polymer} (KK), who argued that chain fluctuations merely renormalize $A$ by a blob with dimension $D$ with $\frac{D}{A}e$ charge. The physical that emerges is that fluctuations do not invalidate the qualitative aspect
of the OSF result. Rather, they effectively reduce the direct
distances between charges so as to renormalize the bare parameters $l_0$, $A$, charge per blob, as was first recognized
by KK. In this limit, local fluctuations inside the blobs are
strong enough to reduce significantly the direct distance between
two consecutive charges by the factor $(\frac{l_0l_B}{A^2})^{\frac{1}{3}}$
which is much less than one. This results in much stronger
effective Coulomb repulsion between two consecutive
blobs. Since the length scale of $l_p$  is much larger than the
blob size $D$ and ordering as implied in these cases does not
refer to the local structure of blobs, we can expect that the
local fluctuations inside a blob not to affect significantly the
property of the chain much beyond the length scale $D$. As a result, we obtain
qualitatively similar dependence of $l_p$ on $\lD$ as that obtained
for stiff polyions. Note that this is relevant only for the
asymptotic case where each blob contains a large number of
segments which roughly obey Gaussian statistics.  The variational theories also predict interesting crossover behavior in which the electrostatic persistence length $\sim \lD^y$ where the exponent $y \le 1$.

\subsection{The linear and sub-linear dependence of $l_e$ on $\lD$}
The OSF theory has formed the basis for interpreting both experiments and simulations. Indeed, a number of experiments have confirmed the quadratic dependence of the electrostatic persistence length $l_e$ on the Debye length $\lD$~\cite{SF1977,WLC4,Koen,Caliskan05}. However, many other experiments~\cite{tricot,Reed,plengthssdna1,ssDNAPincus,Saleh2, ToanJPCM06,ToanMacro2010} and simulations~\cite{Kremer96,Kremer97} have shown that the dependence of $l_e$ on $\lD$ is  linear or even sub-linear. The reason for this discrepancy seems to be because the OSF theory was originally derived for intrinsically stiff polyelectrolytes, those that have the intrinsic persistence length much larger than separation of charges on the backbone $\lo\gg A$, such as dsDNA and carboxymethylcellulose (CMC)~\cite{SF1977}. In the latter experiments, intrinsically more flexible polyelectrolytes, such as ssDNA and RNA were examined. Thus, chain fluctuations or finite size effects can indeed produce deviations from OSF theory. The deviations from OSF behavior have lead to other approaches to study flexible polyelectrolytes.

Barrat and Joanny~\cite{BJ93}, on the other hand, used a variational method in which they replaced the actual Hamiltonian with a ``trial'' one $H_t$, and minimized the resulting free energy $F_{var} = \la H_t\ra - TS_t$, where the average is taking with the Boltzmann weight $\exp\left[-\beta H_t\right]$ with $\beta$ being the inverse of the thermal energy $k_BT$. Here $H_t$ is that of a neutral chain under uniform tension. For flexible chains, they also found that the total persistence length scales as $\lD$. Dobrynin~\cite{Dobrynin2005} also modeled a polyelectrolyte as a wormlike chain with electrostatic interactions and evaluated the bending angle fluctuations in the frame work of the Gaussian variational principle. For semiflexible chain and strongly charged flexible chain, $l_e \sim \lD$, whereas the dependence is sublinear for weakly charged chains.
Ha and Thirumalai (HT)~\cite{Thirum95} used a self-consistent variational theory to calculate the total persistence length of a polyion. The theory is general for both flexible and stiff chains. They found that
\begin{eqnarray}
l_p \sim
\begin{cases}
l_o+l_{OSF}, & \text{if } l_{OSF}\ll l_o\\
\left(l_o\omega_c\right)^{1/2}\lD, & \text{if } l_{OSF}\gg l_o
\end{cases}
\end{eqnarray}
Therefore, for stiff chains ($l_o \gg l_{OSF}$) the electrostatic persistence length $l_e$ is $l_{OSF} \sim \lD^2$, whereas for flexible chains ($l_{OSF}\gg l_o$), $l_e$ is not much different from $l_p$ and scales as $\lD^1$.

The effect of valence in driving collapse of RNA was vividly illustrated using small angle X-ray scattering (SAXS) experiments on a $\sim$200 nucleotide \textit{Azoarcus} ribozyme~\cite{Caliskan05}. From the measured scattering intensity they obtained the distance distribution, $P(r)$. The dependence of the square of the radius of gyration, which corresponds to the second moment of $P(r)$, depends sensitively on the valence of the counterion (see Fig.~\ref{fig:4.1}). The
ribozyme is extended at low cation concentrations and is
compact at elevated values of the counterion concentration
[Fig.~\ref{fig:4.1}]. The collapse transition is highly cooperative in
Mg$^{2+}$ and is much less so in Na$^+$ [Fig.~\ref{fig:4.1}].   It was found that $P(r)$ (Fig.~\ref{fig:4.2}) could be well fit using the asymptotic form for the end-to-end distribution for WLC chains~\cite{Hyeon06JCP}.  The experiments showed the persistence length of this RNA changes dramatically from about 3nm at low ionic concentration to about 1nm at high salt concentrations. More importantly, For both Na$^+$ and Mg$^{2+}$, the persistence
length scales as
$\lD^2$, which is consistent with the OSF theory. From this finding, we find that the
intrinsic persistence length of RNA is $\sim$ 1nm.

The evidence of the linear dependency of the electrostatic persistence length on $\lD$ is provided in a number of experimental studies. Tricot~\cite{tricot} analyzed the intrinsic viscosity-molecular weight dependence of a number of polyelectrolytes such as carboxymethylcellulose in solutions with $I$ ranging from 0.005M to 1.0M. The linear dependency is  found for other types of charged polymers, whose total persistence length is on the order of 10nm or less (see Fig.~\ref{fig:4.4}). Tinland et al.~\cite{plengthssdna1} measured the self-diffusion coefficient of ssDNA fragments using florescence recovery after photobleaching to infer the persistence length. The fitted total persistence length measured in Angstroms is found to be $l_p = 6.42 10^{-8} + 4 I^{-1/2}$ with $I$ measured in molars. Perhaps, the most convincing evidence for the linear variation is found in recent single molecule stretching experiments. Saleh et al.~\cite{ssDNAPincus} combined single molecule stretching data of ssDNA at very low forces with scaling arguments to obtained $l_e\sim I^{-0.51\pm 0.04}$ or $l_e\sim I^{-0.40\pm 0.04}$. The similar behavior is once again reproduced using a general property of the force extension curve at point where the relative extension is about $1/2$ by Toan et al.~\cite{ToanMacro2010}. Toan et al.\cite{ToanJPCM06} also found the linear dependency for synthetic RNA by fitting the force extension curve data by Seol et al.~\cite{Koen} using the Thick chain model~\cite{BiophysJ2005,ToanJPCM06}.

\section{Polyampholytes}

In this section, we present a simple physical model for describing a polyampholyte (PA) chain, especially the weakly-charged case, with a random distribution of opposite charges, in an electrolyte solution (e.g., NaCl dissolved in water).  Here, we do not attempt to elaborate on the known results in the literature~\cite{Joanny91,Rubinstein04}.    There has been much progress in understanding the physical properties of a PA chain under various conditions.  For instance, the shape of a PA chain has been studied extensively for varying solvent quality (good vs poor solvent) and for different charge distributions (e.g., a varying degree of excess charge) (see Refs.~\cite{Joanny91,Rubinstein04} and references therein).  However, the earlier effort has been focused on the low-salt and high-salt limits~\cite{Joanny91,Rubinstein04}.  Our main motivation here is to offer a unified picture, in which both limits are integrated coherently.  In  appropriate limits, our description reproduces known results.   Also, in our approach, the crossover between the two limits is captured in a more transparent manner.    This effort will be beneficial for further illustrating various competing effects in shaping a PA chain and for offering a more coherent picture of such a system.

\subsection{Charge correlations and screening}

Much of our discussion will rely on the concept of ``screening,'' which is well-understood for a simple electrolyte (e.g., NaCl dissolved in water)~\cite{LandauBook}.  Select any charge  and place it at the origin.  Since the entire system is electrically neutral, the charge at the origin will be surrounded and shielded by the ionic cloud of the opposite charge.  If the energy of this system tends to shrink the ionic cloud~\cite{endnote_energy}, the entropy opposes this tendency.  This means that the thickness of the ionic cloud depends on the ion concentration and the temperature~\cite{LandauBook}.   This idea can be extended to PA charges.   While both species ({\i.e.}, PA charges and salt ions) contribute to screening, their roles in shaping the PA chain are opposite.  Along this line, it proves useful to distinguish between ``self-screening'' (screening of a PA charge by other PA charges) and salt-screening ({\i.e.}, screening of PA charges by salts).  Here self-screening refers to the tendency of opposite charges on the chain to be spatially correlated.   This effect is responsible for the electrostatic compaction of a PA chain (under the right conditions) and thus competes with the latter effect, since the surrounding salt ions tend to diminish this tendency~\cite{Joanny91,Rubinstein04}; there is a competition for the spatial correlation of a PA charge with salt ions.  %Because of this, it has proven to be useful to focus on the low-salt and hight-salt limits~\cite{Joanny91,Rubinstein04}.
Despite the difference between self-screening and salt-screening, the electrostatic free energy of a PA chain can be obtained by treating both kinds of charges as forming a Debye-H\"uckel solution (or an electrolyte at the low electrostatic coupling limit~\cite{endnote_low_coupling}).  This is most evident within the theoretical framework known as the random phase approximation~\cite{Joanny93,Rubinstein95}. The level of approximations assumed for this is similar to that for the Debye-H\"uckel approach.  At this level, chain connectivity does not influence the electrostatic free energy of such a (randomly-charged) PA chain (for a given chain size)~\cite{Joanny93,Rubinstein95}. Obviously, non-electrostatic terms are different for the PA chain and for the surrounding electrolyte.

Consider a PA chain consisting of $N$ monomers of size $b$ each, which is inside an imaginary volume $V$ of radius $R$.  The PA chain is assumed to be weakly charged. Let $f_\pm$ be the fraction of positively and negatively charged groups.  Then $f_\pm \ll 1$.  For a neutral chain, $f_+=f_-$ ($\equiv f$).   Let $n_\pm$ be the total density of positive and negative ions whether they are free or on the chain; let $\varepsilon$ be dielectric constant of the solvent %, $\varepsilon_0$ the permittivity of fee space,
and $k_BT$ the thermal energy. % It proves useful to define $\varepsilon=\varepsilon_0 \varepsilon_r$.
The Bjerrum length, at which the Coulomb energy of two electronic charges is equal to the thermal energy $k_BT$, is then expressed as $l_B = e^2/ \varepsilon k_BT$. (Here we adopt Gaussian units as in Refs.~\cite{Joanny91,Rubinstein04}.)  Finally, the Debye screening length, $\kappa^{-1}$, is given by the relation $\kappa^2 = 4 \pi l_B (n_+ + n_-)$.  Outside $V$, $n_\pm$ tends to a constant, $n_0$ ($2n_0$ is the salt concentration at infinity).  Inside $V$, however, the salt ion concentration can be perturbed by the PA charges.   Nevertheless, we %assume that the salt ion concentration remains the same inside $V$.  Accordingly, we
write as $\kappa^2 = \kappa^2_\text{PA} + \kappa_0^2$ inside $V$, where $\kappa_\text{PA}^2 \sim N f l_B /R^3$ and $\kappa_0^2 = 8 \pi l_B n_0$.  Our DH approach will reproduce known results based on a more elaborated approach, in which the salt concentration inside $V$ is determined more systematically~\cite{Joanny91}.

The electric potential due to a point charge $e$ is given by
\begin{equation}
\Psi ({\bf r}) ={e\over  \varepsilon} {e^{-\kappa r}\over r}
.\end{equation}
This is a potential screened by both PA charges and salt ions.  To isolate the contribution of PA charges to screening, one has to subtract the salt contribution as follows
\begin{eqnarray}
\Psi_\text{PA} &=& \lim_{r \rightarrow 0} \left[ \Psi (r) - {e \over \varepsilon} {e^{-\kappa_0 r}\over r}\right] \nonumber \\
&=& - {e \over \varepsilon} (\kappa -\kappa_0).
%&=& - {1\over 2}l_B \left( \sqrt{ \kappa_\text{PA}^2+ \kappa_0^2} -\kappa_0 \right)
%&=& - {e \over \epsilon} \left(\sqrt{\kappa_\text{PA}^2+\kappa_0^2} -\kappa_0 \right)
\label{eq:Psi}
\end{eqnarray}
The electrostatic energy gain per charge due to self-screening or the polarization free energy of the PA chain is
\begin{eqnarray}\label{eq:E_PA}
{E_\text{PA} \over k_BT} = {1\over 2} {e \Psi_\text{PA} \over k_BT} %&=& - {1\over 2}l_B (\kappa - \kappa_0) \nonumber \\
&=& - {1\over 2}l_B \left( \sqrt{ \kappa_\text{PA}^2+ \kappa_0^2} -\kappa_0 \right)
.\end{eqnarray}
The factor $1/2$ is to avoid double counting of interaction pairs. The total polarization free energy of the PA chain can be written as
\begin{equation}
{F_\text{PA}\over k_BT} \sim -N f l_B \left( \sqrt{{N f l_B\over R^3} + \kappa_0^2} -\kappa_0 \right)
\label{eq:freePA}
,\end{equation}
where $\kappa_\text{PA}$ is expressed explicitly in terms of PA parameters.  Note here that the electrostatic free energy differs from the energy contribution by a numerical prefactor.   Clearly, this free energy favors chain collapse.

\subsection{Flory theory: a $\Theta$-solvent case}

In Flory theory, the non-electrostatic free energy can be expanded in powers of the monomer density $\rho = N/R^3$ as
\begin{equation}
{F   \over k_BT} \sim \left( v b^3 \rho^2 + w b^6 \rho^3  \right) R^3
\label{eq:nonelecF}
,\end{equation}
where $v$ and $w$ are the second and third virial coefficient, respectively~\cite{deGennes}.  The total free energy is the sum of $F_\text{PA}$ and $F$:
\begin{equation}
{F_\text{total} \over k_BT} \sim \left( v b^3 \rho^2 + w b^6 \rho^3  \right) R^3 -N f l_B \left( \sqrt{{N f l_B\over R^3} + \kappa_0^2} -\kappa_0 \right)
.\end{equation}
Here we mainly focus on a $\theta$ solvent in which $v=0$, even though our general approach can readily be extended to other cases (see Refs.~\cite{Joanny91,Rubinstein04}).   For $v=0$, we show our approach reduces to existing ones in low salt and high salt limits.
In this case, the two-body term, {\i.e.}, the first term in Eq.~\ref{eq:nonelecF}, vanishes.   The equilibrium size of a PA chain is then determined by the balance between $F_\text{PA}$ and the three body term, {\i.e.}, the second term in Eq.~\ref{eq:nonelecF}~\cite{Joanny91}. The equilibrium $R$ satisfies
\begin{equation}
\left( Nfl_B \over R^3\right)^2 \left( {Nf l_B \over R^3} + \kappa_0^2\right)^{-1/2} - w b^6 \left({N \over R^3} \right)^3 =0
\label{eq:4.6}
.\end{equation}
Because of the functional form of this relation, one can easily find that $R\sim N^{1/3}$.   The dependence of $R$ on other parameters is more involved.  After analyzing the relation in Eq.~\ref{eq:4.6} in the two limiting cases, $\kappa_0 \rightarrow 0$ and $\kappa_0 \rightarrow \infty$, we will present our numerical solution for the intermediate range of $\kappa_0$.  For simplicity, we set $w=1$.

{\it Low-salt limit: $\kappa \rightarrow 0$}.  In this case, Eq.~\ref{eq:4.6} reproduces the known result~\cite{Joanny91}
\begin{equation}
R \sim N^{1/3} b \left(  {b \over f l_B} \right)^{1/3}
\label{eq:Rlowsalt}
.\end{equation}
The chain size decreases as $f$ increases, as expected based on the following picture.  For larger $f$, the effect of self-screening or the PA effect is more pronounced and shrinks the chain size more effectively.

{\it High-salt limit: $\kappa \rightarrow \infty$}.   In this case, Eq.~\ref{eq:4.6} reduces to
\begin{equation}
\left( Nfl_B \over R^3\right)^2 {1\over \kappa_0} - b^6 \left({N \over R^3} \right)^3 =0
.\end{equation}
This reproduces the known result for the high-salt limit~\cite{Joanny91}
\begin{equation}
R \sim N^{1/3} b \left({\kappa_0 b^3 \over f^2 l_B^2} \right)^{\frac{1}{3}}
\label{eq:Rhighsalt}
.\end{equation}
The chain is swollen by added salts.  This finding is consistent with our earlier assertion that self-screening competes with salt-screening.

{\it Crossover}.  At the crossover between the two limits, the PA size expressed in Eq.~\ref{eq:Rlowsalt} becomes  comparable to that in Eq.~\ref{eq:Rhighsalt}.  This leads to the following crossover condition~\cite{Joanny91}
\begin{equation}
{f l_B \over b^2} \sim \kappa_0
.\end{equation}
In fact, the term on the left hand side is the reciprocal of the self-screening length, $\kappa_\text{PA}$, for a PA in the low-salt limit:
\begin{equation}
\kappa_\text{PA} \sim \sqrt{{N f l_B \over R^3}} \sim {f l_B  \over b^2}
.\end{equation}
The crossover condition thus reads $\kappa_\text{PA} \sim \kappa_0$.

For the intermediate salt range, we have solved Eq.~\ref{eq:4.6} numerically and plotted our results in Fig.~\ref{fig:4.6}.  Our results in Fig.~\ref{fig:4.6} show how added salts expand the PA chain.  A key to understanding this behavior is that the tendency for PA charges to be correlated is weakened by the surrounding salt ions, since they are equally correlated with the salt ions.  This competition underlies the $\kappa_0$ dependence of $R$ shown in Fig.~\ref{fig:4.6}.   \\

In summary, our DH approach combined with Flory theory offers a unified framework for describing the equilibrium properties of a polyampholyte solution in the presence or absence of added salts.  It not only reproduces known results in the low-salt and high-salt limits but also shows the crossover between the two limits.    In the next section, we complement this approach using the blob picture of a PA chain.

\subsection{Blob picture of a polyampholyte chain}

The notion of blobs has been proven to be useful for illustrating various aspects of polymer systems~\cite{deGennes}. The section is devoted to presenting a blob picture of a PA solution, offering a ``visual'' guide to the approach presented in earlier subsections.  For a weakly charged PA chain, which we consider here, the chain statistics at short length scales will not be perturbed by the PA effect.  Within this length scale, the PA chain resembles an ideal chain or a random walk (in a $\theta$ solvent).  Beyond this, the PA effect will collapse the chain.  This means that there exists a special length, denoted as $\xi_\text{PA}$, that separates these two regimes.  To obtain this length, consider a section of the chain, consisting of $g$ monomers.  For sufficiently small $g$ ($>1$), the size of this section is given by $r \sim b g^{1/2}$.  The free energy of this section is dominated by the entropic term $r^2/g b^2$.  As $g$ increases, the PA term becomes more important.  To set up the crossover condition, find $g$ at which the two terms are equally important:
\begin{equation}
\left({r^2 \over g b^2} \right)_{r \sim b g^{1/2}}\sim  g f l_B \left( \sqrt{{g f l_B \over r^3} + \kappa_0^2} -\kappa_0 \right)_{r \sim b g^{1/2}}
.\end{equation}
This leads to
\begin{equation}
g f l_B \left( \sqrt{{f l_B\over g^{1/2}b^3}+\kappa_0^2} -\kappa_0\right) \sim 1
\label{eq:g}
.\end{equation}
The $g$ value that satisfies Eq.~\ref{eq:g} is the minimum number of steps the random walk has to take until it starts to feel the PA effect.  The blob size is then $\xi_\text{PA} \sim b g^{1/2}$.  For $N \gg g$, the chain can be viewed as a compact stack of many such blobs.   This means that $R \sim \xi_\text{PA} (N/g)^{1/3} \sim N^{1/3} b g^{1/6}$.

{\it Low-salt limit}.   As $\kappa_0 \rightarrow 0$, Eq.~\ref{eq:g} results in $g \sim b^2/(f l_B)^2$ and $\xi_\text{PA} \sim b^2/f l_B \sim \kappa_{PA}^{-1}$, consistent with Ref.~\cite{Joanny91}.

{\it High-salt limit}.  For large $\kappa_0$, Eq.~\ref{eq:g} reduces to $g\sim \kappa_0^2 b^6/(f l_B)^4$. The length equivalent is then $\xi_\text{PA} \sim \kappa_0 b^4/(f l_B)^2 \sim \kappa_0/\kappa_\text{PA}^2$, in agreement with Ref.~\cite{Joanny91}.

In both limits, this blob picture reproduces the results for $R$ in Eqs.~\ref{eq:Rlowsalt} and \ref{eq:Rhighsalt} obtained by free energy minimization.  Also one can show that $F_\text{PA} \sim k_BT \times (N/g)$.  This is simply the free energy cost for ``redirecting'' the random walk so as to fill the space compactly.  Hence $k_BT$ per blob.

{\it Intermediate salt concentration}.  For the intermediate range of $\kappa_0$, $g$ has been obtained numerically and plotted in Fig.~\ref{fig:4.6} (see the dotted lines) for a few choices of $f$; we have used the same color scheme as for the $R$-$\kappa_0$ plot.  As shown in the figure, $\xi_\text{PA}$ grows in magnitude as $\kappa_0$ increases.  This is paralleled with the earlier finding that the PA chain is swollen by salt.

{\it Surface tension}.  Our DH approach to a PA chain is based on the assumption that $N$ is arbitrarily large.  In this case, the ``surface effect'' is minimal, as long as the PA chain is overall spherical.  Under different conditions, however, this picture may break down~\cite{Rubinstein04}. For instance, if stretched by an external force or a net-charge repulsion, a PA chain will break into many smaller subunits.  An important contribution to the PA free energy arises from the fact that PA charges on the periphery (the surface of the chain) are less effectively screened by other PA charges than those inside.   This unfavorable free energy cost per unit area is called the surface tension, denoted as $\gamma$.  %In the case of a DH electrolyte, the surface tension was formulated by allowing a spatially-varying $\kappa$
Following the scaling approach adopted in Ref.~\cite{Rubinstein04}, $\gamma$ is the number of blobs on the surface per unit area and is readily given by
\begin{equation}
{\gamma \over k_BT} \sim {1\over \xi_\text{PA}^2} \sim \left\{ \begin{array}{ll} {\left(f l_B \right)^2 /b^2}, & \mbox{low salt}\\
{\left( f l_B \right)^4 / \left( \kappa_0 b^4 \right)^2},  & \mbox{high salt}
\end{array}
\right.
\label{eq:gamma}
.\end{equation}
Finally, we have plotted the surface tension $\gamma$ as a function of $\kappa_0$ in Fig.~\ref{fig:4.6} (see the inset).   Since $\gamma$ is a direct result of PA effects, its magnitude is diminished as $\kappa_0$ increases, as shown in the figure.

While the notion of the surface tension will be useful in some contexts (e.g., the formation of a necklace globule~\cite{Rubinstein04}), we only focus on overall spherical PA's here.  An excess charge on a PA chain can drive a conformational transition to a pearl-necklace structure from a more spherical globule (referred to as a polyelectrolyte (PE) effect).  This may be realized at a low-salt limit.  It is worth noting that the low-salt limit here is different from that for the PA effect, since the PE effect is longer-ranged.  Salt-screening is not felt sensitively by the PA effect, as long as $\kappa_\text{PA} \sim f l_B/b^2 \gtrsim \kappa_0$, which is $N$-independent.  On the other hand, the PE effect will be screened unless $R \kappa_0 \lesssim 1$ or $\kappa_0 \lesssim 1/R$.  For the long-chain case, this imposes a very strict condition on salt concentrations. In conclusion, the low-salt limit for the PA effect can be realized in a wide range of salt concentrations, while that for the PE effect is prohibitively narrow, as long as $N \gg 1$.  We thus focus on overall ``spherical'' PA's, whether they are collapsed or swollen (e.g., self-avoiding walk chains).

\section{ Polyampholytes with excess charges}

 For several reasons, PA chains can carry an unbalanced, excess charge.  Recall $f_\pm$ is the fraction of positive and negative charges on the PA chain.  Then the chain will carry a net charge unless $f_+ = f_-$.  The net charge repulsion now enters into the picture and competes with the PA effect.  To focus on this polyelectrolyte effect, smear out all the charges on a PA chain, which is assumed to be inside an imaginary volume $V$.  The excess electrostatic energy stored in this volume is expressed as
\begin{eqnarray}
{E_\text{PE} \over k_BT} &=& {l_B \over 2} (f_+ - f_-)^2 \rho^2  \int\!\int_{{\bf r},  {\bf r}' \in V} {e^{-\kappa |{\bf r}-{\bf r}'|} \over |{\bf r}-{\bf r}'|} d {\bf r} d {\bf r}' \nonumber \\
&\sim & V l_B (f_+ - f_-)^2 \rho^2 \left( {1\over {\bf k}^2+\kappa^2}\right)_{{\bf k}=0} \nonumber \\
&\sim& {l_B (f_+ - f_-)^2 \over \kappa^2} \rho^2 R^3
.\end{eqnarray}
Here the subscript of $\kappa_0$ was dropped for simplicity; ${\bf k}$ is the Fourier conjugate to ${\bf r}-{\bf r}'$.  Only ${\bf k}=0$ contributes to $E_\text{PE}$, since the monomer density is assumed to be uniform inside $V$.  The PE free energy thus scales as
\begin{equation}
{F_\text{PE}\over k_BT} \sim {l_B (f_+ - f_-)^2 \over \kappa^2} \rho^2 R^3
\label{eq:freePE}
.\end{equation}
This estimate is relevant for the case $R \gg \kappa^{-1}$.  It is tempting to relate this to the two-body term in Eq.~\ref{eq:nonelecF} and interpret the PE term as renormalizing the second virial coefficient $v$, as in Ref.~\cite{Joanny91}.  Strictly speaking, this reasoning is valid only if each ``chain segment'' remains spherical, in the sense that its width is equal to or comparable with its length.  A crucial concept that describes the PE effect on chain shape is the electrostatic persistence length, denoted here as $l_e$, which was originally introduced over three decades ago~\cite{Odijk1977,SF1977}.  While this concept has been widely used in the literature, ironically, there has been a controversy over whether $l_e \sim \kappa^{-2}$, as originally predicted~\cite{Dobrynin2009,Thirum99,Fixman10} (also see references therein).

With this subtlety in mind, let's insist on using the PE term in Eq.~\ref{eq:freePE} and compare it with  the PA term in Eq.~\ref{eq:freePA}.  These two effects are comparable if $F_\text{PE} \sim |F_\text{PA}|$.  For the case $\kappa_0 > \kappa_\text{PA}$, this implies that
\begin{equation}
\left( f_+ + f_-\right)^2 l_B \sim \left( f_+-f_-\right)^2 \kappa^{-1}
\label{eq:PEvsPA}
.\end{equation}
(One can arrive at this by considering Eq.~(4.11) in Ref.~\cite{Joanny91}).  For a random PA chain, $f_+=f_-$ and $|f_+ -f_-| \sim 1/\sqrt{N}$.  Despite some uncertainty in the $\kappa$ dependence of Eq.~\ref{eq:PEvsPA}, it is clear that the PE effect can be easily dominated by the PA effect.  For different solvent chemistry or polymerization processes ({\i.e.}, $f_+ \ne f_-$), however, the PE term can be dominant and will expand the chain.  As a result,  $R \sim N^\nu$, where $\nu \approx 3/5$ if $R \kappa  \gg 1$ or $\nu =1$ if $R \kappa \lesssim 1$.  We will not attempt to refine earlier results for PA chains with excess charges, especially in regard to the $\kappa$ dependence of their size.    As more results become available, they can be incorporated into our approach.

\section{Elastic response of flexible polyelectrolytes}
The response of flexible polyelectrolytes such as ssDNA and RNA to mechanical force has further given a fundamental understanding of elasticity of PEs and has further clarified the changes in persistence length as the ionic concentration is varied. In fact, double stranded DNA molecules, which we have seen are intrinsically stiff polyelectrolytes, have long been the major polymers of interest in single molecule stretching experiments~\cite{busta,WLC5,WLC1,WLC3}. For dsDNA and related stiff PEs, a wormlike chain (WLC) model readily explains  the measured force-extension data~\cite{WLC5,WLC1,WLC3} as well as the distribution of the end-to-end distance~\cite{Valle2005}. For flexible polyelectrolytes, however, it has been difficult to account for the data using polymer models alone, thus raising the possibility that microscopic structures might matter. For these systems, it is probably the strong interplay between the electrostatic effects and the small intrinsic stiffness of the molecules that complicates the physics of stretching. For ssDNA and RNA in particular,  base pairing between pairs of nucleotides along the chain backbone could be an additional complicating factor. Here, we will just focus on the situations in which the base pairing interactions are negligible.

We begin with a survey of experiments to indicate the diversity of responses of charged flexible PEs. Dessinges et al.~\cite{dessinges} used magnetic tweezers to obtain force-extension curve (FEC) for a 11 kilo base ssDNA molecules, in the range of force from 0.05pN to about 50pN. In order to extract structural information of the molecules, the authors used an extensible freely-jointed chain with electrostatic interactions. The electrostatics is modeled using the Debye-H\"uckel potential with an effective charge density, $\nu$, along the backbone first described by Zhang et al.~\cite{YangZhang}
\begin{eqnarray}
E_{elec.} = {\nu^2\over \varepsilon}\int ds_i ds_j {\exp\left(-{r_{ij}\over\lD}\right)\over r_{ij}}
\end{eqnarray}
where $\varepsilon = 80$ is the dielectric constant of water, $r_{ij}$ is the spatial distance between two points $i$ and $j$, and the double integral is taken along the chain contour. The force-extension data for this model is determined through Monte Carlo simulations. It appears that the model reproduces well stretching data in several different ionic conditions. In particular, in solution of 10mM phosphate buffer with $\lD = 1.87$nm, where the base pairing interactions are suppressed, the FEC of the ssDNA appears to be almost straight for at least two decades of force in the log-linear scales (see Fig. 4 of ref.~\cite{dessinges}. At the values of $\nu = 1.28e/$nm and Kuhn length of 1.6 nm, the theoretical FEC  follows very  the experimental data. The extensible wormlike chain and the extensible freely jointed chain models without electrostatic interactions can only fit the FEC for forces at least 20pN or even higher. Even so the persistence length obtained with the extensible wormlike chain model is too small, $l_p = 0.21$nm. (It should be noted that although the FEC portion for $f\ge 20$pN was plotted in the same figure, the data had actually been obtained previously for overstretched dsDNA in a solution at much higher ionic concentration of 150mM Na$^+$ by Rief et al.~\cite{gaub99a}.)

In 2004, Seol et al.~\cite{Koen} built synthetic RNA constructs made of only uracils, or poly(U) (no possibility for base-pair interactions) and examined their elastic properties using optical tweezers.  The solutions considered have the ionic strength ranging from 5mM to 500mM, and the force range is from about 0.5pN to 50pN. The authors used, instead,  a modified extensible WLC model with electrostatic interactions where the relative extension is given by~\cite{WLC1,Barra93}
\begin{eqnarray}
x = 1 - \int {dq\over 2\pi } {1\over l_p q^2 + f/K_BT} + {f\over S}
\end{eqnarray}
where $S$ is the stretch modulus in unit of force (pN) and the scale dependent persistence length is $l_p = \l0+\lo\til\kappa(q)$. Thus, it is assumed {\em a priori} that the electrostatic persistence length is  given by the OSF theory. The modified FJC model with electrostatic interactions could not fit the data well at high ionic concentrations. However, WLC model appears to reproduce the experimental FEC at high concentration of Na$^+$ down to about 10mM. The resulting persistence length does not depend on the force scale for concentration up to 500mM, whereas it decreases by two fold from $f=1$pN to about 50pN.

More recently, Saleh et al.~\cite{ssDNAPincus,Saleh2} used magnetic tweezers to stretch ssDNA molecules that are specially treated to avoid base pairing even at the lowest forces. The ionic concentrations were varied over a broad range (20mM to 5000mM of Na$^+$), and the minimum force is about 0.08pN, whereas the maximum force is similar to those in the aforementioned works. Instead of modeling electrostatic through the Debye-H\"uckel potential, they use the blob picture and scaling arguments~\cite{pincus,deGennes} for self-avoiding chains. Besides, demonstrating for the first time the Pincus regime $x\sim f^{2/3}$ at salt concentrations up to 2000mM, the data also shows the logarithmic dependence of the extension on the force, as observed earlier~\cite{YangZhang,dessinges,Koen}. The two regimes are demarcated by a crossover force $f_c$ at a characteristic  extension $x_c$. When all the FECs at different salt concentrations are normalized by $f_c$ and $x_c$, they collapse fairly well onto a single master curve. The scaling arguments show that $f_c\sim k_BT/l$ and $x_c\sim (v/l^3){1/3}$, where $l$ is the Kuhn length, which is twice  the total persistence length, and $v$ is the excluded volume between the Kuhn segments. Since $f_c$ and $x_c$ can be extracted from the FECs, $v$ and $l$ then can be estimated without any assumption on their relation. As a result, $l \sim I^{-0.51\pm 0.4}$, which is almost close to $I^{-0.5}$ or the Debye length $\lD$. In addition, at $I=3000$mM, there is effectively no excluded volume effect due to the counterion condensation (see next section) that renders the excluded volume $v$ negligibly small ($\Theta-$ condition). Under this condition,  the FEC can then be fitted very well using the standard WLC model~\cite{WLC1} but not with the FJC model. The extracted persistence length $l_p$ is $0.62\pm 0.01$ nm, which can be treated as the ``bare'' persistence length of ssDNA molecules. The seemingly diverse responses of a single chain can be explained using a unified theory (see below).

In 2006, Toan et al.~\cite{ToanJPCM06} used their Thick Chain model, which views any polymer as a tube with  uniform thickness, to fit the FEC data in ref.~\cite{Koen}. The excluded volume effects along the chain backbone leads to an effective persistence length. The extracted parameters are the effective monomer length and the polymer thickness, both of which can be combined in a simple formula to produce the persistence length. Besides producing a good fit to the data for almost all salt conditions, the model shows that the effective thickness of RNA varies linearly with the Debye length. More importantly, it has been shown the resulting persistence length can be fitted with $l_p = \l0 + c \lD$, where $\l0\approx 0.66$nm and $c = 0.43$ is a  constant. Thus, in contrast to the results by Yeol et al.~\cite{Koen}, which used the WLC model with electrostatic interactions that assumed a OSF type persistence length, the dependence here is linear or somewhat sub-linear.

In a recent paper published in 2010, Toan and Thirumalai~\cite{ToanMacro2010} using a combination of geometrical arguments and scaling theory to derive a model-free unified theory for semiflexible polymer stretching at high force. The theory can be used for polyelectrolytes in moderate to high salt concentrations. One of the main results of the theory is that the apparent elasticity of a polymer is inherently force-dependent. The FEC will appear to be that of a WLC model in the force range of $\fl = k_BT/l_p$ to $\fh \approx 4 k_BT l_p/b^2$ with $b$ being the monomer length, and the polymer behaves as a FJC for forces higher than $\fh$. Although similar results for WLC-like models were obtained previously~\cite{Kierfeld,Livadaru,angelo}, the results are general and hold good for any semiflexible chain regardless of the details of the monomer-monomer interactions. The numerical values of the two crossover forces for typical ssDNA and RNA molecules are $\fh \approx 4$pN and $\fh\approx 50$pN, which turn out to be relevant to the reported discrepancy in the elastic behaviors of structurally similar molecules, ssDNA and RNA. The higher crossover force is about the maximum force than can be achieved in most of the experiments~\cite{dessinges,Koen,ssDNAPincus}, which means that the WLC behavior is more likely to be observed. This is also true for the data cited in ref.~\cite{dessinges}, where the FJC fits well the data above 50pN (in fact the data is from another work~\cite{gaub99a} done at different ionic concentration (see above). In fact, fitting the same high force data using a generalized formula for both the WLC and FJC regime, Toan et al.~\cite{ToanMacro2010} show that the fit is reasonably good and produce a physically meaningful persistence length of $0.72$nm.
Moreover, the theory naturally leads to a simple, called the $1/2$-rule, which states that the extension $x$ at force $f=\fl=k_BT/l_p$ is $1/2$. That means the persistence length could be quickly estimated from the FEC without doing any fit, by calculating the force at $x=1/2$. Indeed, the rule does lead to the estimate of the persistence lengths of RNA~\cite{Koen} and ssDNA~\cite{ssDNAPincus} in different ionic solutions, that are similar to each other and also scale linearly in the Debye length (see Fig.~\ref{fig:4.5}).

\renewcommand{\sectionmark}[1]{\markright{\MakeUppercase{\thesection\ \ \ Simulations of collapse of an...}}{}}
\section{Simulations of collapse of an isolated  polyelectrolyte chain}
A theoretical description of PE collapse is relevant in a number of applications including RNA folding and conformational fluctuations of disordered proteins.  The initial event in the folding of RNA is counterion driven collapse. The kinetics of collapse of this process is driven by a number of factors such as solvent quality, valence, shape, and size of counterions.  Although the final structure is unaltered the pathways in the coil-globule transition is dependent on the details of interactions between ions and RNA. In addition, DNA collapse into toroidal structures can also mediated by counterions. These considerations make the study of collapse of PEs (and the related PAs) important.

Lee and Thirumalai~\cite{Lee2001} analyzed the collapse of flexible polyelectrolytes in poor solvents using simulations. In their model, each monomer is approximated as a van der Waals sphere with radius $b/2$ that carries a charge of $-e$.  Two successive monomers are connected by a spring with spring constant $3k_BT/b^2$. The polyelectrolyte chain is  in a box with counter ions, whose valence varies from $+1$ to $+4$. The number of the ions is such that the total system is neutral. The non-electrostatic interaction between the particles (monomers or counter-ions) $i$ and $j$ with radii $\sigma_i$ and $\sigma_j$ respectively that are at a spatial distance $r_{ij}$ is modeled using Lennard-Jones potential
\begin{eqnarray}
{H_{\text{LJ}}\left(r_{ij}\right)} = \epsilon_{\text{LJ}}\left[\left({r_0 \over r_{ij}}\right)^{12} - 2 \left({r_0 \over r_{ij}}\right)^{6}\right]
\end{eqnarray}
where $r_0 = (\sigma_i+\sigma_j)/2$. The parameter $\epsilon_{\text{LJ}}$ is used to control the quality of the solvent, which is expressed in terms of the second virial coefficient $v_2$ roughly equal to the excluded volume
\begin{eqnarray}\label{eq:v2}
v_2 = \int_v \text{d}^3 \rv \left(1 - \text{e}^{-\beta H_{\text{LJ}}} \right).
\end{eqnarray}
For poor solvents or hydrophobic $v_2 < 0$, and the strength of the hydrophobicity depends on the value of $v_2$ (see Fig.~\ref{fig:4.7}). The number of monomers considered was $N=240$, for which the effective $\Theta$-temperature corresponds to $\epsilon_{\text{LJ}} = 0.5$ and $v_2 = -1.9b^3$. The Coulomb potential is used to model interactions between the charged particles $i$ and $j$ with valences $z_i$ and $z_j$ respectively
\begin{eqnarray}
{H_C\left(r_{ij}\right)\over k_BT} = {l_B z_i z_j\over r}
\end{eqnarray}
The chain dynamics is  followed using Brownian Dynamics simulations. The time unit is $\tau=b^2\zeta/2k_BT$, where $\zeta$ is the friction coefficient of a monomer.

\subsection{Effect of Valence and Solvent Quality on Collapse Dynamics}
The dynamics of collapse of the PE chain is determined by a balance between hydrophobic interactions and charge renormalized electrostatic potentials. The chain would be in extended conformation if the electrostatic repulsion is dominant. On the other hand, the valence-dependence effective electrostatic attractions between monomers due to the Manning condensation (of counterions onto the chain) could make the chain collapse. In particular, let us consider the case of near $\Theta$ and  hydrophobic conditions.

In near $\Theta$ solvent ($\epsilon_{\text{LJ}} = 0.3$ and $v_2 = -0.06b^3$), the collapse kinetics is monitored through the normalized radius of gyration $R_g(t)/R_c$, where $R_c = 4.08b$ is the size of compact globule. The collapse is strongly dependent on $z$, and occurs readily for $z=3$ and $4$ within the time of $150\tau$ from initial condition. For $z=2$, the collapse only happens after $600\tau$, whereas it is not seen for $z=1$ even up to the time limit of the simulation. There are two effects of counterion condensation: (i) the overall charge of the polyanion is greatly reduced and (ii) there is an effective attractions due to the excess charges which effectively makes the solvent poor and gives rise to the collapse. In the hydrophobic condition $v_2\le -10.3 b^3$, on the other hand, the neutral chain would adopt a compact conformation. Thus the counterion condensation would further accelerate the collapse. In deed, with $\epsilon_{\text{LJ}} = 2.0$ and $v_2 = -25.8b^3$, the collapse already occurs at $t=200\tau$ for $z=2$ and at about $600\tau$ for $z=1$.

The collapse is also found to happen in three stages. The first stage corresponds to the condensation of the counterions to the charged backbone, that is driven mostly by diffusion and occurs on the order of $25\tau$ independently of the valance $z$. The second stage is the formation of pearl-necklace structures of globular clusters containing both monomers and counterions. The clusters are mainly local, namely the monomers in a cluster are predominantly neighbors, and are connected by strings. In the third stage, all the clusters merge and the largest cluster grows at the expense of smaller ones (see below for a theoretical explanation).

The collapse mechanisms depend on the valency of the counterions. For $z=1$, the monovalent counterions when condensed combine with the charges on the backbone to form random dipoles of magnitude $p\approx eb$. When the attraction between two dipole exceeds $k_BT$, i.e. the spatial distance is less than $\lb$, contacts between distinct segments of chain can form. Because this is a very short distance, the globular clusters are predominantly local and their sizes are very small while the number of them is large. Thus it takes very long time to reach the global compact structure. When the counterions are multivalent, they can both neutralize the negative charge on the monomer and provide excess charge of $|(z-1)e|$. Bare monomer charges that are separated by a large distance along the contour can be attracted to the positive charge in a process called ``ion-bridging''. The range over which such attractive interactions are effective increases with $z$. Thus the size of the initial clusters and the efficiency of the collapse also increase with $z$.

\subsection{Phase diagram in the $(v_2,z)$ plane}
The observations concerning the structure of the collapsed
globules together with simple scaling arguments can be
used to construct a valence-dependent phase diagram
for strongly charged polyelectrolytes in poor solvents
($v_2 < 0$). When counterion
condensation takes place, the decrease in the effective
charge of the polyanion can be computed by equating
the chemical potential of the free and condensed counterions
(two-phase approximation). Generalizing the
arguments of Schiessel and Pincus~\cite{Schiessel} to arbitrary $z$, the authors
find that the total charge of the polyelectrolyte decreases
from $Nf_e$ ($f_e$ is the fraction of charge, which is 1 here) to $N\til{f_e}\approx k (L/\lb)(1/z)$
where $k \approx -\ln\phi$ and $\phi$ is the volume fraction of the free
counterions. The size of the polyelectrolyte is given by
$L\approx k^2b^2N/\lB z^2$ provided $\lB > k^2z^{-2}bN^{1/2}$. As the quality
of the solvent decreases to a level such that thermal blob
size $\xi_T \approx b^4/|v_2| < \xi_{el} \approx \lB z^2/k^2$ (size of the electrostatic
blob) then the chain condenses to a globule. The boundary
dividing the stretched and collapsed conformation
is obtained by equating $\xi_{el}$ and $\xi_T$ and is given by $|v_2| \approx
b^4k^2/\lB z^2$.

In the globular phase, two regimes are found, one
corresponding to the Wigner crystal and the other a
Wigner glass. The boundary between the two is obtained
by equating the gain in the energy upon condensation
($\approx ze^2/d\varepsilon$) to the attractive interaction due to the poor
solvent quality ($kTv^2b^{-6}$). This leads to the condition $|v_2|\approx
z^{1/2}b^{5/2}\lB^{1/2}$. To determine the boundary more precisely
we have to account for the induced attraction between
monomers after counterion condensation. Thus, the
actual determination of the boundary separating the
Wigner crystalline regime and the glassy regime requires
equating $ze^2/d\varepsilon$ and $kTv_{2R}^2b^{-6}$ where $v_{2R}$ is the
renormalized second virial coefficient, and $d$ is approximately the distance between the condensed counterion and the backbone charge. This argument shows that the shape of the PE chain requires,
in a nontrivial way, the coupling of electrostatic interactions
and effects coming from solvent quality. In light
of this, the boundary indicated in Fig.~\ref{fig:4.7} should be
regarded as qualitative.

The boundary between the two collapsed phases is
difficult to determine from the simulation because of the
extreme slow dynamics in the glassy phase. For the purpose
of illustrating the phase diagram, we assume that the
ionic glass is an equilibrium phase. To validate the
phase diagram, extensive simulations were performed for
all $z$ $(=1, 2, 3, 4)$ at values of  $\lb^{-1}$ and $|v_2|$ indicated
by asterisks in Fig.~\ref{fig:4.9} The structural conformations are in
qualitative accord with the predicted phase diagram.

\renewcommand{\sectionmark}[1]{\markright{\MakeUppercase{\thesection\ \ \ #1}}{}}
\section{Theory of Collapse Dynamics}
Although there are a number of theoretical models that describe the collapse kinetics of homopolymers with applications to protein folding there is only one study that has considered the rate of PA and PE collapse. The theory proposed by Lee and Thirumalai~\cite{Lee00JCP} (LT) adopts an approach developed by Pitard and Orland~\cite{Pitard98EuroPhysLett} to describe uncharged homopolymer collapse in poor solvents. For simplicity, we consider the collapse of PAs, which as will argue bears a strong resemblance to the behavior expected for PEs as well.

A model system, which may captures
some generic aspects of biomolecules, is a polyampholyte, which is a linear polymer chain that has both positive
and negative charges along the backbone. The very large muscle protein titin associated with sarcomere contains regions that have PA characteristics as do many intrinsically disordered proteins. Because of the
simultaneous presence of positive and negative charges the
shape of PAs is determined by competition between the two
conflicting interactions. The repulsion between the
charges tends to swell the chain (the polyelectrolyte effect)
while the attraction (polyampholyte effect) tends to collapse
the chain. It is known that if the total charge, $Q$,  of PA is less
than $Q_c \sim \pm Ne$ ($N$ is the number of monomers) the chain adopts a globular
conformation, and is extended otherwise. If $Q > Q_c$ then it can be shown, using an analogy to Rayleigh instability of charged droplets, that the chain can be visualized in terms of pearl-necklace structures, which is similar to that found in weakly charged PEs. The LT theory showed that  collapse to a global globular structure, which can only take place occur when $Q < Q_c$, occurs in two major steps.  In the first stage the
chain forms a metastable pearl-necklace structure reminiscent
of the equilibrium structures for $Q > Q_c$. In the second
stage the pearls (domains) merge leading to the compact
globular conformation. The theory for the intermediate time
regime (when pearl-necklace structures form) was developed
by adopting the procedure suggested by Pitard and Orland~\cite{Pitard98EuroPhysLett}. The LT theory showed that the square of the radius of gyration decays as
\begin{equation}
R_g^2(t) \approx R_g^2(0)(1 - (\frac{t}{\tau_{PA}}))^{\alpha}
\end{equation}
where $\alpha = \frac{5}{4}$ and the characteristic time $\tau_{PA} \sim N^{\frac{1}{\alpha}}$. On the time scale $\tau_{PA}$ metastable pearl-necklace structures (local structures connected by strings) form. Interestingly, formation of such structures have also been observed in the coil-globule transition in uncharged homopolymers.

\textit{Dynamics of PE collapse:} The theory described for PA
can qualitatively predict the collapse dynamics
of strongly charged PEs. Let a strongly charged PE (fraction
of charged monomers is close to unity) chain be initially in the $\Theta$-
solvent with respect to the neutral polymer. At $t=0$, we
imagine a quench to low enough temperatures so that counterions
start to condense.  Upon condensation of counterions
the conformation of the chain approximately resembles
that of a PA. The relevant length scales for PE are the electrostatic
blob length  $D \approx \frac{l_Bz^2}{k^2}$ and $\xi_T \approx a_0 \frac{\theta}{\theta - T}$
where z is the valence of counterions, $k = -ln \phi$ ($\phi$ is the
volume fraction of free counterions), and  $\Theta$ is the collapse
temperature of uncharged polymer. If $\xi_T$ is not too small, the
PE chain undergoes a sequence of structural changes
en route to the collapsed conformation.  After counterion
condensation PE evolves toward a metastable pearl-necklace
structure.  For weakly charged polyelectrolytes such structures
are the equilibrium conformations in poor solvents.
The dynamics of this process can be described using the theory developed to describe collapse of PA.  We assume that
shortly following the quench to low temperature the counterions
condense onto the PE chain. The time scale for this
process is diffusion limited, with values that are much smaller than time in
which the macromolecule relaxes. Upon condensation of a
multivalent cation ($z \ge 2$), the effective charge around the
monomer becomes ($(z-1)e$). If the locations of the divalent
cation are random and if the correlations between the counterions
are negligible then in the early stages, a PE chain
with condensed counterions may be mapped onto an evolving
random PA.  With this analogy we suggest that
the pearl-necklace-like structures that are found in the collapse
of strongly charged PEs  should also be governed by Eq. (35).

\textit{Late stages of collapse:}  At long times the pearl-necklace structures
merge  to form compact collapsed
structures. This occurs by the largest cluster growing
at the expense of smaller ones, which is
reminiscent of the Lifshitz–Slyozov mechanism~\cite{Lifshitz61JPhysChemSol}. If this analogy is correct then we expect
the size of the largest cluster $S(t)$ to grow as $S(t) \sim t^{\frac{1}{3}}$.
The collapse is complete when $S(t) \sim R_g(t \rightarrow \infty) \approx N^{\frac{1}{3}}$ which implies that the characteristic collapse time
is  $\sim N$.

Langevin
simulations for strongly charged flexible PEs where the
collapse is induced by multivalent counterions have shown that late stage coarsening indeed occurs by the Lifshitz-Slyozov growth mechanism. Figure 1 shows a representative example
of the conformations that are sampled in the dynamics
of approach to the globular state under $\Theta$-solvent conditions.
Both panels show that in the later stages of collapse the
largest clusters in the necklace-globule grow and the smaller
ones evaporate. This lends support to the proposed Lifshitz-
Slyozov~\cite{Lifshitz61JPhysChemSol} mechanism.
To estimate , we have used simulations to calculate the
number of particles that belong to the largest cluster $N_S(t)$ as
a function of time. Figure 2 shows the linear increase of
of the number of particles $N_S(t)$ at long times from which the growth time increases linearly with $N$ as expected from the LS mechanism. . The change of slope for long times is due to the finite
size effects and indicates the completion of the globule formation.

The mechanism of approach to the globular
state for PA and PE should be similar. The collapsed
conformations are reached via metastable pearl-necklace
structures. For PE the driving force for forming such structures
is the counterion mediated attractions. Charge fluctuations
in PA lead to pearl-necklace structures. The lifetime of
pearl-necklace structures depends on energy barriers separating
merged and unmerged clusters. The necklace-globule
conformation consists of $n$ globules with nearly vanishing
net charge that are in local equilibrium. The free energy of
the $i^{th}$ globule is $F_i \sim (\frac{4 \pi}{3})\Delta F R_i^3 + 4 \pi \sigma R_i^2$ where $R_i$ is
the radius of the $i^{th}$ globule and $\sigma$ is the surface tension. Note that $\Delta f$ is the same before and after the merger of
clusters. The free energy difference between a conformation
consisting of two clusters and one in which they are merged
is $\Delta F \sim (8 \pi \sigma - 4(2)^{\frac{2}{3}} \pi \sigma) (\frac{N}{n})^{\frac{2}{3}}$.  Charge fluctuation in
each globule is $q_0 (N/n)^{1/2}$. If the Coulomb energy fluctuation
of each globule ($\delta E \sim (N/n) q_0^2/a_0 (\frac{N}{n})^{\frac{2}{3}}$)
free energy difference between the conformation with two
separated clusters and the one where they are merged, then
the system spontaneously grows to a large cluster.

\section{Conclusions}
We have only described a few topics dealing with properties of isolated polyelectrolytes and polyampholytes with some applications to biological systems. There are a variety of issues especially such as the viscosity of PEs that are puzzling and warrant explanation. In addition, bundling of charged systems (actin and microtubule for example) is another topic with applications in biology that have not been adequately described.  Only recently, polyelectrolyte concepts have been applied to understand conformational fluctuations in intrinsically disordered proteins~\cite{Schuler10PNAS}. Detailed theoretical and experimental studies are needed to clarify the role that polyelectrolyte effects and ion-PE interactions play both in the synthetic and biological world.

\section{Acknowledgement}
This work was supported by a grants from NSERC (Canada) to BYH and the National Science Foundation to NMT and DT.

\bibliography{bookchapter_bib_v3}

\begin{mcitethebibliography}{54}
\bibitem[end()]{endnote_low_coupling}
Crudely speaking, the low electrostatic coupling means that the interaction
  between charges is weaker than the thermal energy $k_BT$. More precisely, the
  electric energy of a PA chain per charge in Eq.~\ref{eq:E_PA} (and the energy
  of the surrounding salt ions per ion) should be smaller in magnitude than
  $k_BT$. In the low-salt limit, one can find $\kappa_\text{PA} \ell_B < 1$.
  This is the condition required for a weakly-charged PA chain.\relax
\mciteBstWouldAddEndPunctfalse
\mciteSetBstMidEndSepPunct{\mcitedefaultmidpunct}
{}{\mcitedefaultseppunct}\relax
\EndOfBibitem
\end{mcitethebibliography}

\newpage
{\bf \Large Figure captions}\\

\begin{figure}[h]
  \centering
  \includegraphics[width=4in]{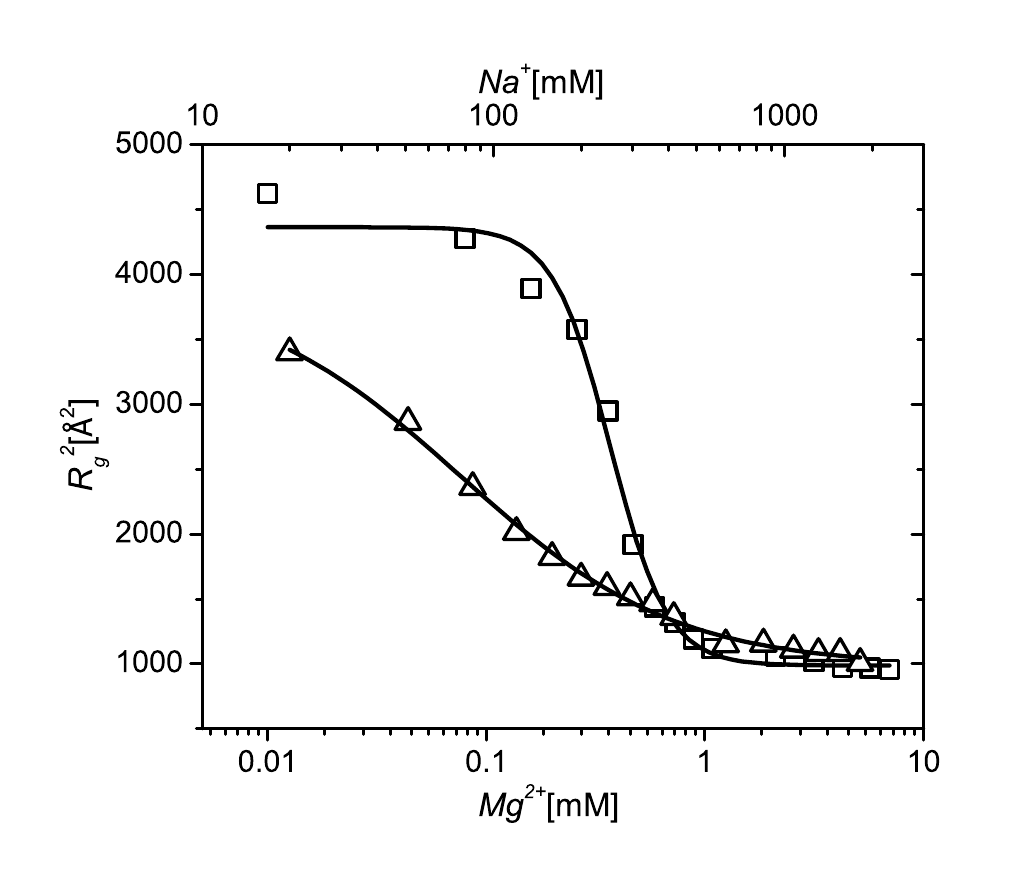}
  \caption{}
  \label{fig:4.1}
\end{figure}

{\bf Figure~\ref{fig:4.1}}: (Reproduced from~\cite{Caliskan05}) The dependance of $R_g^2$ on Mg$^{2+}$ (squares) and Na$^+$ (triangles) concentration. Solid lines are the fits using Hill equation in the form $A\{1-[\text{Mg}^{2+}]^n/ \left(C_m^n+[\text{Mg}^{2+}]^n\right) \} +y_0$, where $A$, $n$, and $y_0$ are adjustable parameters. The best fit values for $n$ are 3.33 and 1.20 for Mg$^{2+}$ and Na$^+$, respectively. Clearly, cooperativity of the folding transition increases as the valence of ions increases.

\begin{figure}[h]
  \centering
  \includegraphics[width=4in]{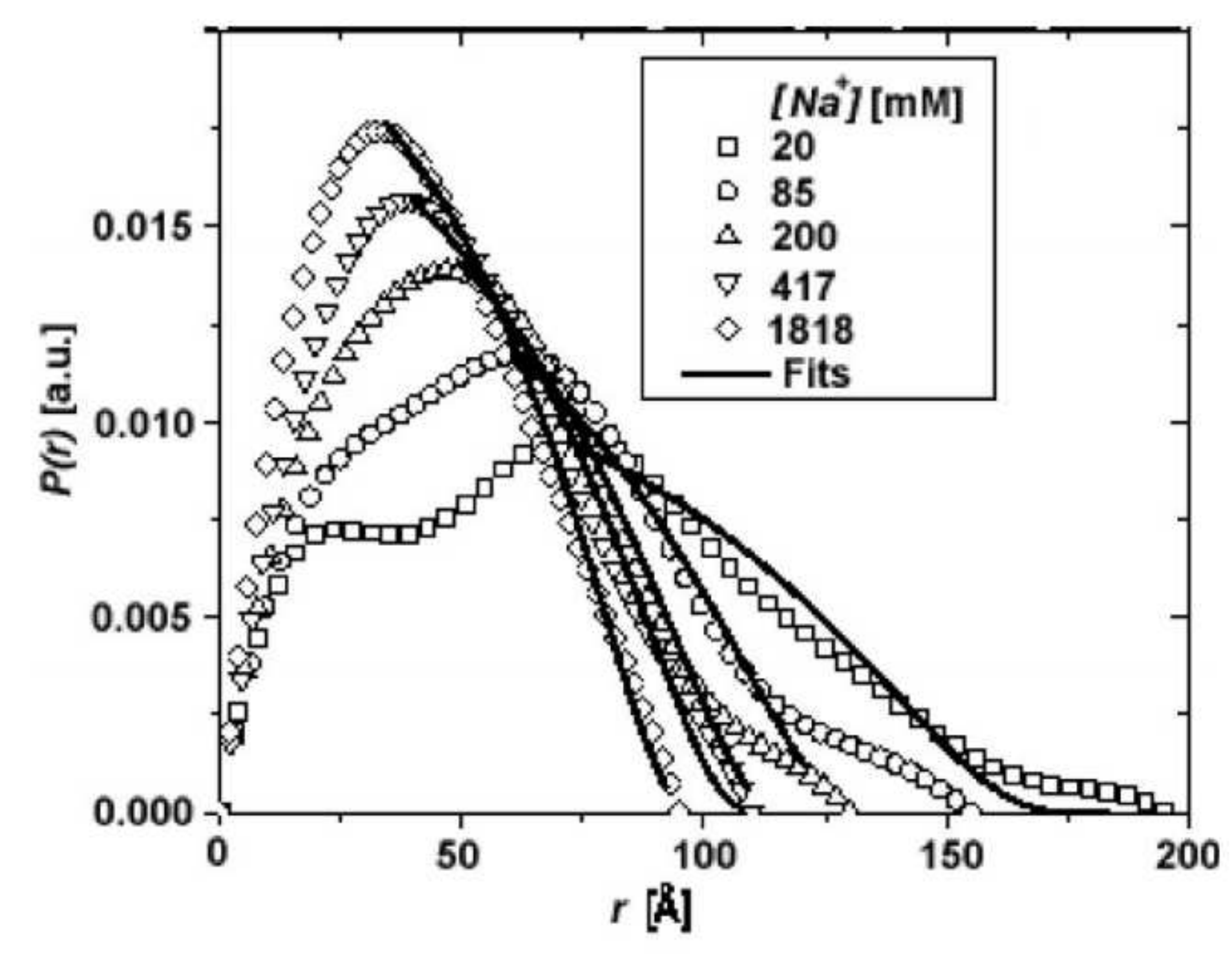}
  \caption{}
  \label{fig:4.2}
\end{figure}

{\bf Figure~\ref{fig:4.2}}: (Reproduced from~\cite{Caliskan05}) Distance distribution functions $P(r)$ at various Na$^+$
concentrations at 32 \textdegree C are obtained by inverting the measured scattering intensity by a Fourier transform. The solid lines are fits of $P(r) \sim \exp\left(-{1\over 1-x^2}\right)$, which is the expected asymptotic result for WLC. The concentrations of counterions are given in the insets.

\begin{figure}[h]
  \centering
  \includegraphics[width=4in]{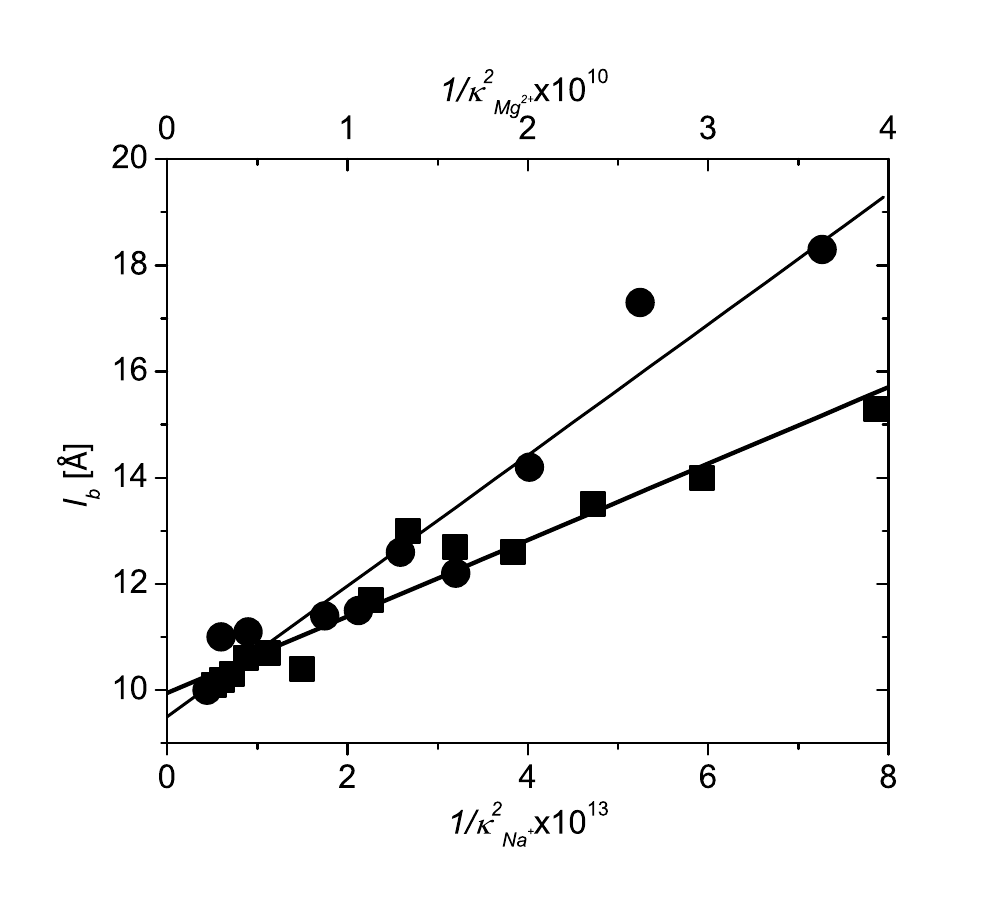}
  \caption{}
  \label{fig:4.3}
\end{figure}

{\bf Figure~\ref{fig:4.3}}: (Reproduced from~\cite{Caliskan05}) Dependence of $l_p$ on $1/\kappa^2$ in Mg$^{2+}$ (solid circles) and Na$^+$ (solid squares). Lines represent fits to the data. Note that  the $1/\kappa^2$ for Mg$^{2+}$ is given on top.

\begin{figure}[h]
  \centering
  \includegraphics[width=4in]{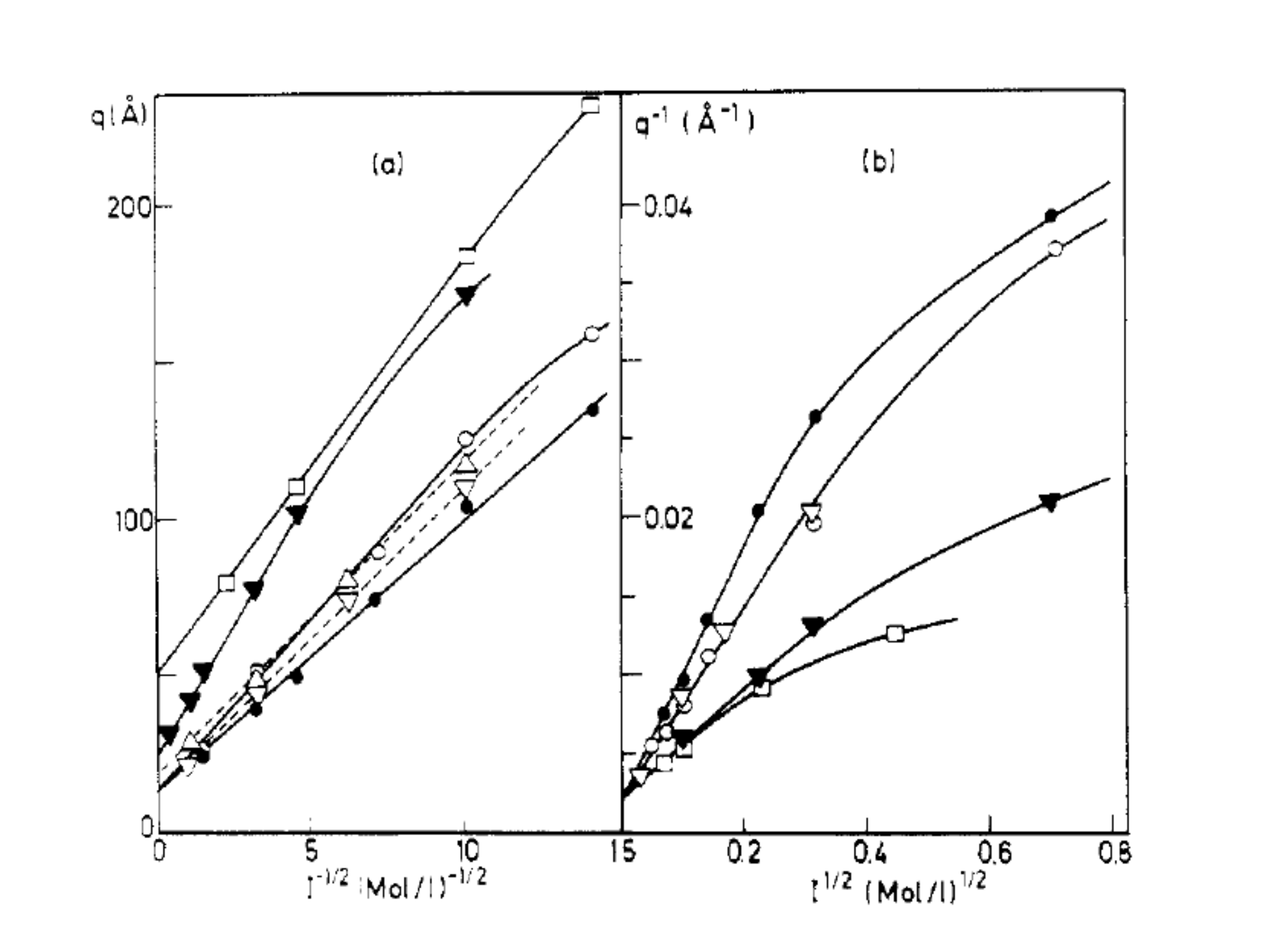}
  \caption{}
  \label{fig:4.4}
\end{figure}

{\bf Figure\ref{fig:4.4}}: (Reproduced from~\cite{tricot}) Variations of the experimental persistence length with the reciprocal of the square root of the ionic strength for sodium poly[((acry1ami- do)methyl) propanesulfonate] ($\blacktriangledown$), sodium (carboxymethy1)cellulose ($\square$), sodium polyacrylate ($\circ$), sodium poly(styrenesu1fonate) ($\bullet$), and the sodium salt of an alternating copolymer of isobutyl vinyl ether and maleic anhydride at two degrees of neutralization $\alpha = 0.5$ ($\triangle$) and at $\alpha = 1.0$ ($\nabla$).

\begin{figure}[h]
  \centering
  \includegraphics[width=4in]{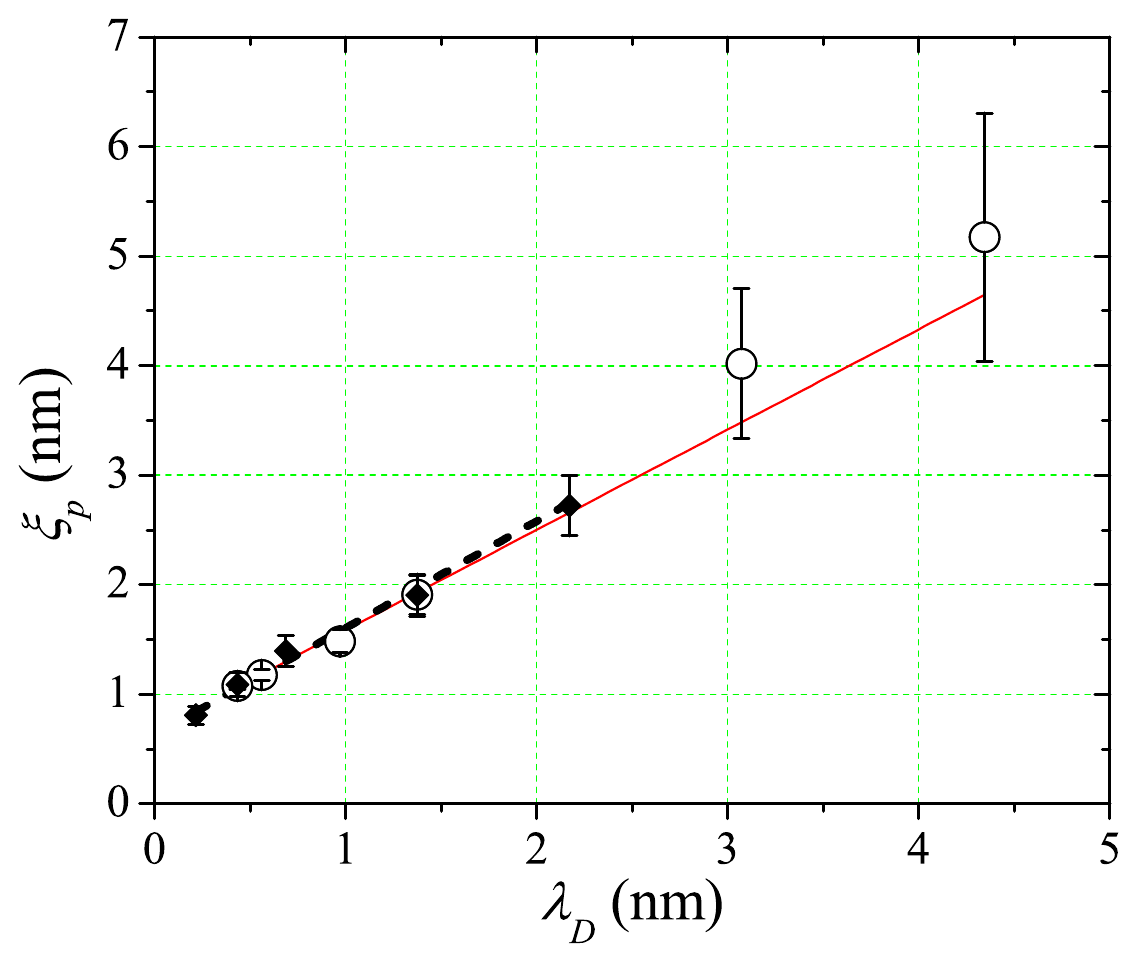}
  \caption{}
  \label{fig:4.5}
\end{figure}

{\bf Figure~\ref{fig:4.5}}: (Reproduced from~\cite{ToanMacro2010}) Analysis of the experimental results. Estimates of $l_p(\lD)$ for poly(U) (circles) and ssDNA (filled symbols) using the $1/2$-rule. Linear fits to the data (solid and dashed lines) yielded the bare persistence lengths $l_p \approx 0.67$nm for poly(U) and $l_p\approx 0.63$nm for ssDNA.

\begin{figure}[h]
  \centering
  \includegraphics[width=4in]{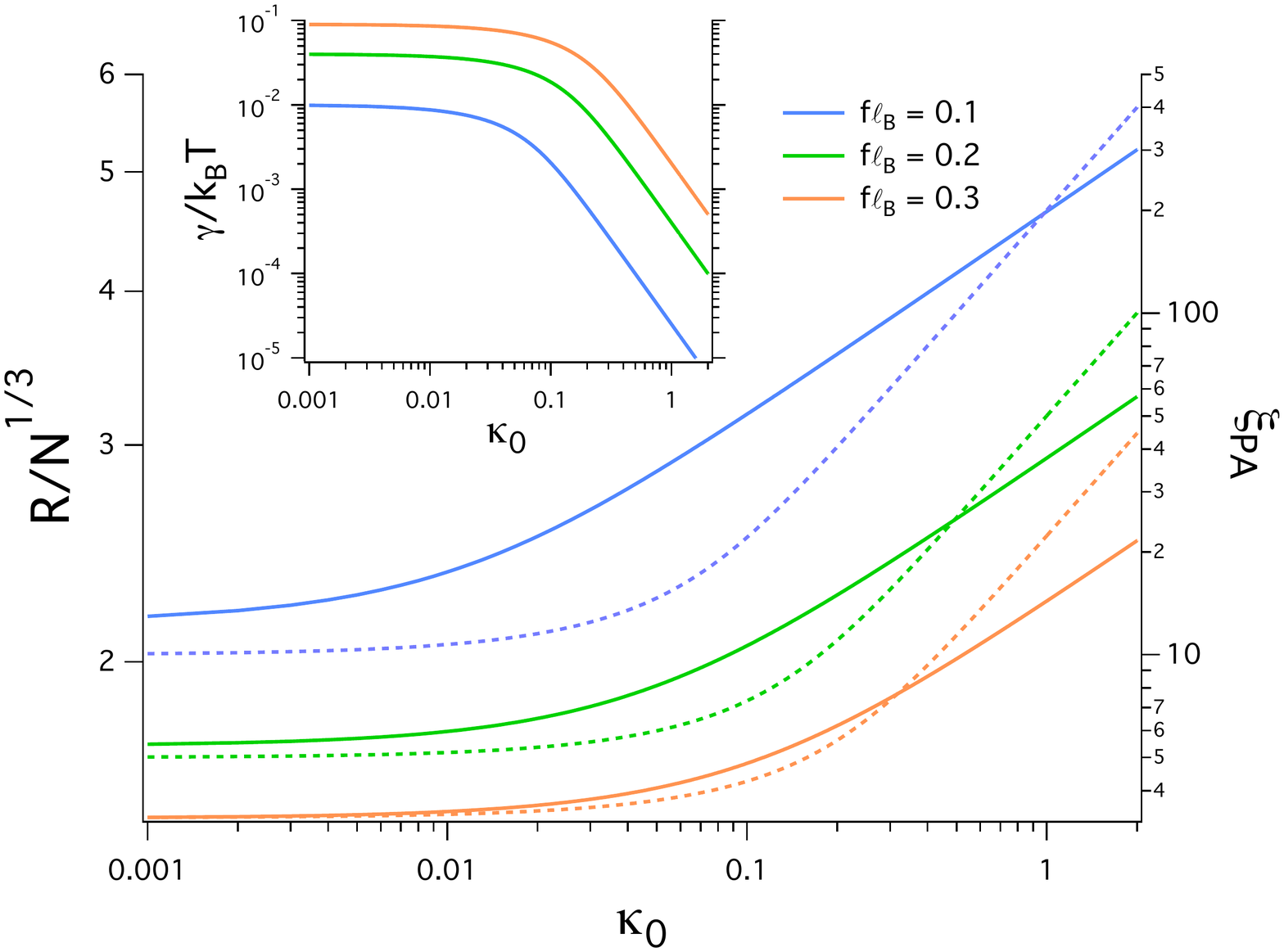}
  \caption{}
  \label{fig:4.6}
\end{figure}

{\bf Figure~\ref{fig:4.6}}:  The equilibrium size of a polyampholyte chain as a function of $\kappa_0$ for a few choices of $f$: $fl_B=0.1, 0.2,0.3$.  Our results show how the PA chain is swollen  by salt-screening, which tends to weaken the polyampholyte effect: the competition for the screening of PA charges by salt ions is more pronounced at high salt concentrations.  Also plotted is the size of polyampholyte blobs, $\xi_\text{PA}$ (the same color scheme used for $R$-$\kappa_0$ graphs is adopted).  Our result for $R$ and $\xi_\text{PA}$ are useful for illustrating the crossover from the low-salt to high-salt limit. Finally, we have plotted the surface tension of a PA chain, $\gamma \sim 1/\xi_\text{PA}^2$ as a function of $\kappa_0$ (see the relevant discussion around Eq.~\ref{eq:gamma}).

\begin{figure}[h]
  \centering
  \includegraphics[width=4in]{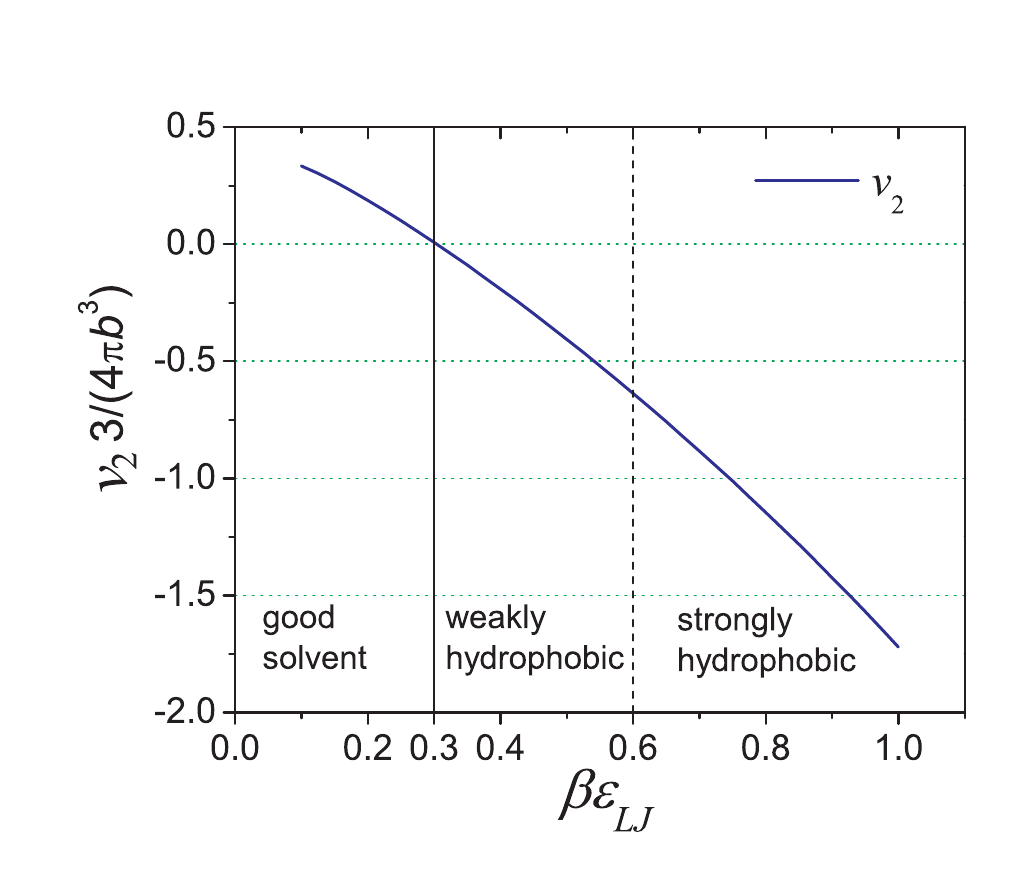}
  \caption{}
  \label{fig:4.7}
\end{figure}

{\bf Figure~\ref{fig:4.7}}: Second virial coefficient as a function of $\epsilon_{\text{LJ}}$ from Eq.~\eqref{eq:v2}. The classification of solvent quality based on the values of $v_2$ are
shown.

\begin{figure}[h]
  \centering
  \includegraphics[width=4in]{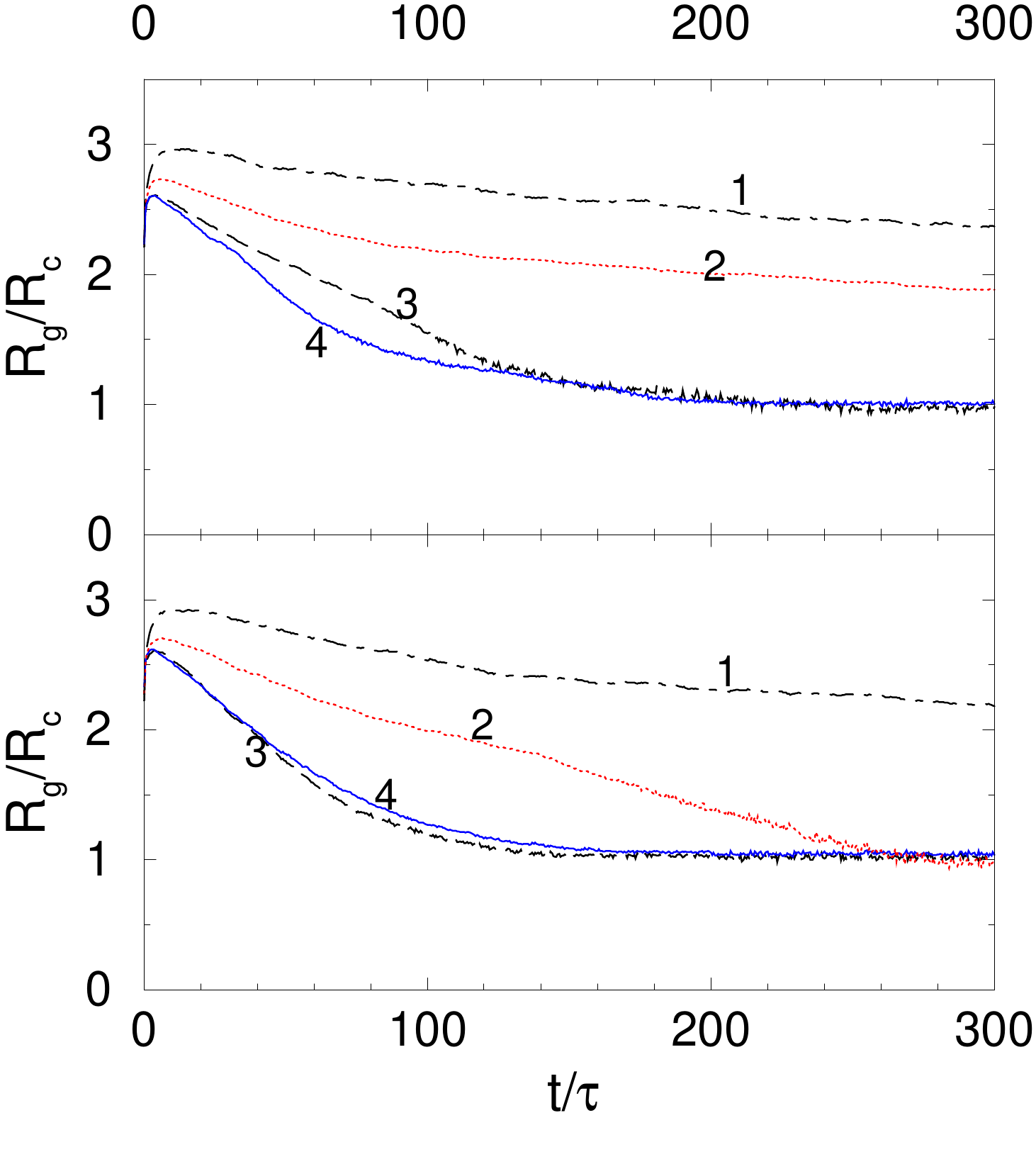}
  \caption{}
  \label{fig:4.8}
\end{figure}

{\bf Figure~\ref{fig:4.8}}: The time dependence of the radius of gyration following a quench from $\Theta$-solvent to poor solvent condition. The value of $N = 240$ and $\lb = 5.3b$. Top panel is for $v_2 = -0.06b^3$ whereas bottom panel is for $v_2 = -3.69b^3$. The numbers on the curves denote the valence of the counterions.

\begin{figure}[h]
  \centering
  \includegraphics[width=4in]{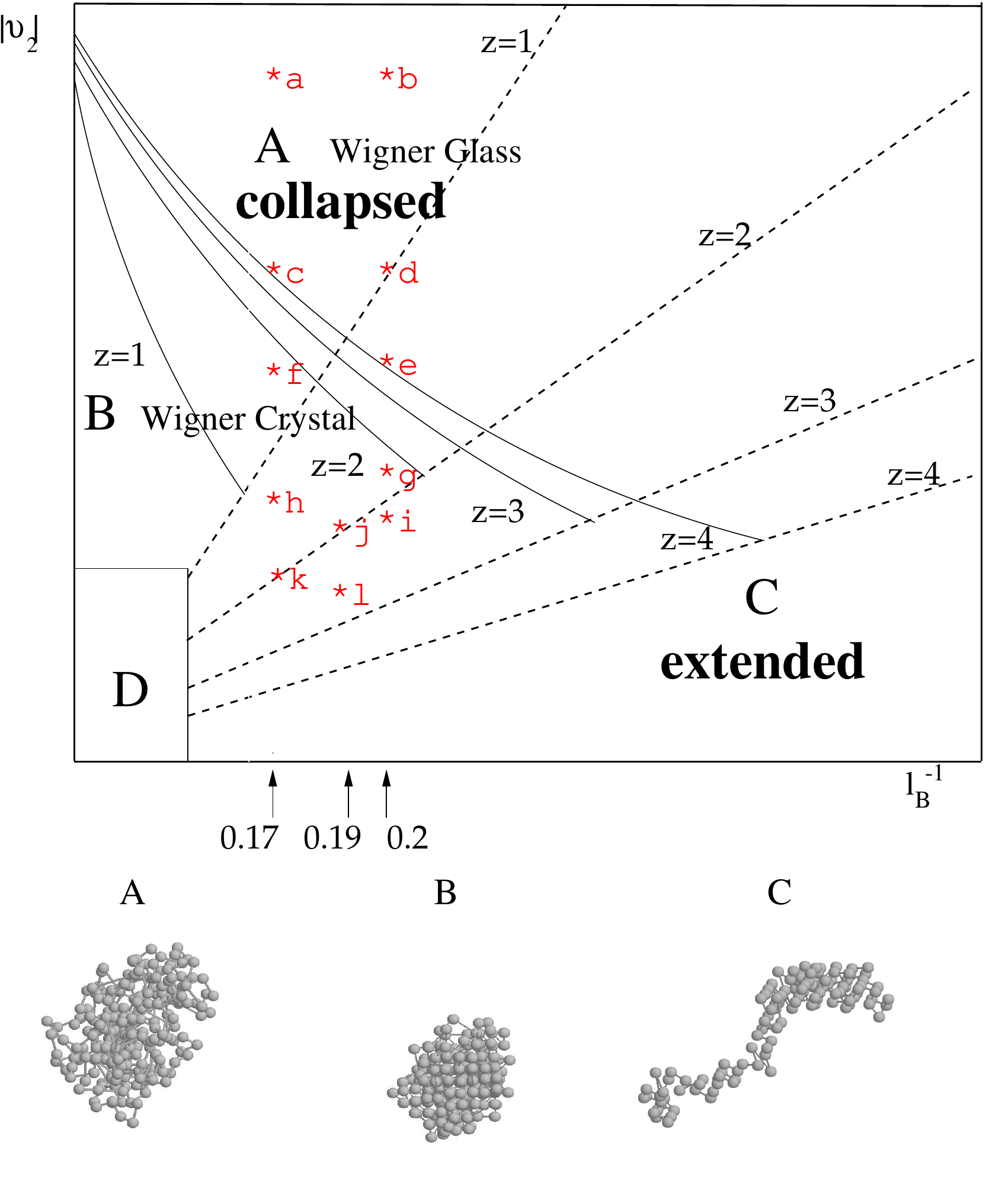}
  \caption{}
  \label{fig:4.9}
\end{figure}

{\bf Figure~\ref{fig:4.9}}: Valence dependent diagram of states in the ($|v_2|$ and $l_B^{-1}$) plane for strongly charged PE. The dashed lines represent the boundary between stretched and collapsed states and depend on $z$. The $z$-dependent solid lines in the collapsed region separate Wigner crystalline region from the Wigner ``glassy'' region. The asterisks are the simulation results with each data point corresponding to four $z$ values ($z$ = 1 - 4). Pictures of the conformation of the chain in region A,B and C are also shown. The $l_B^{-1}$ values are indicated by arrows. The letter (a-l) near the asterisks corresponds to values of $|v_2|/b^3$= 62.34,62.34,25.82,25.82,15.20,12.36,7.51,4.47,3.91,3.69,0.34,0.06 respectively.

\newpage

\newpage

\newpage

\newpage

\newpage

\newpage

\newpage

\newpage

\newpage

\end{document}